\documentclass[sigconf]{acmart}

\AtBeginDocument{%
  }

\usepackage{tabularx} %
\usepackage{hyphenat,balance}
\usepackage[fleqn,tbtags]{mathtools}
\usepackage{contour}
\usepackage{microtype}
\usepackage{multirow}
\usepackage{colortbl}
\usepackage{amsmath}
\usepackage{xspace}
\usepackage{textcomp}
\usepackage{enumitem}
\setitemize{noitemsep,topsep=0pt,parsep=0pt,partopsep=0pt} %
\usepackage{soul}

\graphicspath{{./images/}}

\def \eg {{\emph{e.g.},\thinspace}}
\def \ie {{\emph{i.e.},\thinspace}}

\definecolor{llightgray}{RGB}{230,230,230}

\newcommand{\rev}[1]{{\color{black} #1}}

\newcommand{\revv}[1]{{\color{black} #1}}

\newcommand{\methodName}{\emph{DancingBox}\xspace}

\copyrightyear{2026}
\acmYear{2026}
\setcopyright{cc}
\setcctype{by-nc-nd}
\acmConference[CHI '26]{Proceedings of the 2026 CHI Conference on Human Factors in Computing Systems}{April 13--17, 2026}{Barcelona, Spain}
\acmBooktitle{Proceedings of the 2026 CHI Conference on Human Factors in Computing Systems (CHI '26), April 13--17, 2026, Barcelona, Spain}
\acmPrice{}

\acmSubmissionID{2972}

\begin{document}

\title{\methodName: A Lightweight MoCap System for Character Animation from Physical Proxies}

\author{Haocheng Yuan}
\orcid{0009-0007-1717-1585}
\affiliation{%
 \institution{University of Edinburgh}
 \streetaddress{10 Crichton Street}
 \city{Edinburgh}
 \country{United Kingdom}
 }
\email{H.C.Yuan@ed.ac.uk}

\author{Adrien Bousseau}
\orcid{0000-0002-8003-9575}
\affiliation{%
 \institution{Inria, Universit\'{e} C\^{o}te d'Azur}
 \streetaddress{2004 route des lucioles}
 \city{Valbonne}
 \country{France}
}
\email{adrien.bousseau@inria.fr}

\author{Hao Pan}
\orcid{0000-0003-3628-9777}
\affiliation{%
 \institution{Tsinghua University}
 \streetaddress{No. 30 Shuangqing Road, Haidian District}
 \city{Beijing}
 \country{China}
}
\email{haopan@tsinghua.edu.cn}

\author{Lei Zhong}
\orcid{0000-0003-1778-9282}
\affiliation{%
 \institution{University of Edinburgh}
 \streetaddress{10 Crichton Street}
 \city{Edinburgh}
 \country{United Kingdom}
 }
\email{zhongleilz@icloud.com}

\author{Changjian Li}
\orcid{0000-0003-0448-4957}
\affiliation{%
 \institution{University of Edinburgh}
 \streetaddress{10 Crichton Street}
 \city{Edinburgh}
 \country{United Kingdom}
 }
 \email{chjili2011@gmail.com}

\begin{abstract}

Creating compelling 3D character animations typically requires either expert use of professional software or expensive motion capture systems operated by skilled actors. We present \methodName, a lightweight, vision-based system that makes motion capture accessible to novices by reimagining the process as digital puppetry. Instead of tracking precise human motions, \methodName captures the approximate movements of everyday objects manipulated by users with a single webcam. These coarse proxy motions are then refined into realistic character animations by conditioning a generative motion model on bounding-box representations, enriched with human motion priors learned from large-scale datasets. To overcome the lack of paired proxy–animation data, we synthesize training pairs by converting existing motion capture sequences into proxy representations. A user study demonstrates that \methodName enables intuitive and creative character animation using diverse proxies, from plush toys to bananas, lowering the barrier to entry for novice animators. 

\end{abstract}



\begin{teaserfigure}
  \includegraphics[width=\textwidth]{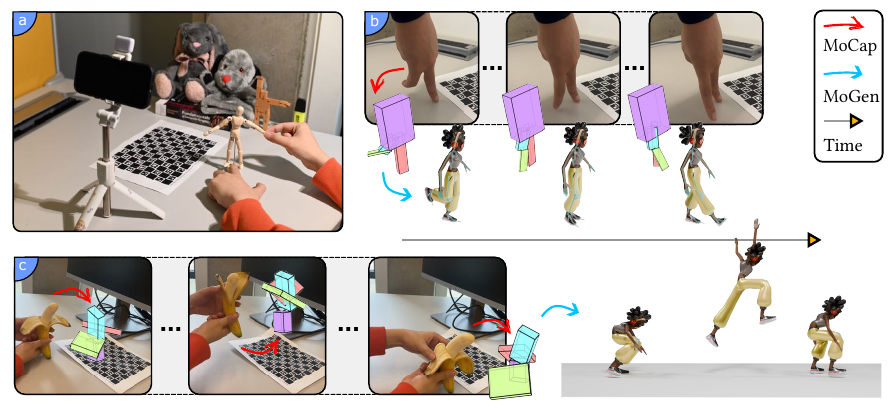}
  \caption{
    a) Overview of our \methodName system. The interactive setup consists of a mounted web camera (\eg a smartphone) facing the scene, where users manipulate a physical proxy to convey their intended character animation, and a planar object (\eg a checkerboard) for calibrating the up direction of the ground.
    b) From the captured monocular video input of finger performance, \methodName first estimates a sequence of bounding boxes using a lightweight vision-based motion capture module (\ie the red arrow, MoCap). It then transforms the coarse box motion into a natural skeletal motion through a conditional motion generation module (\ie the blue arrow, MoGen). 
    c) Our system supports a wide range of physical proxies, such as plush toys, humanoid puppets, or even everyday objects, like a pen or a book. Notably, a banana with peeled skin, mimicking a jumping action with outstretched hands, is faithfully interpreted into the corresponding character motion.  
    For improved visualization, skeletal motions are retargeted to the ‘Michelle’ character from Mixamo. Please see supplemental material for the animation clips of all results shown in the paper.
    }
    \label{fig:teaser}
      \Description{A teaser figure with three blocks - a) the interactive system setup, b) the high-level algorithm pipeline, and c) an appealing example for creating a jumping action.}
\end{teaserfigure}

\maketitle

\section{Introduction}

Creating 3D character animations is at the core of computer graphics, serving widely across films, games, and mixed reality. A typical way of producing character animations is by manually specifying the position and orientation of body parts using professional software, which is known to have a steep learning curve and requires a great time and effort to achieve high-quality animation, even with expertise \cite{zhong2025sketch2anim}. Alternatively, motion capture systems track the body parts of human actors to apply the same motion to virtual characters \cite{desmarais2021review, moeslund2006survey}. But such systems often rely on specialized hardware to achieve high-quality tracking, and only skilled actors are able to perform diverse, compelling motions. 

To make character animation more accessible, researchers in human-computer interaction have explored physical proxies as an interface for users to control character motion~\cite{hung2024fingerpuppet,hung2022puppeteer,ye2020aranimator,li2024anicraft,glauser2016rig}. But these puppetry-based systems 
rely on custom hardware or are limited to specific motion types.

In this paper, we propose a lightweight puppetry-based motion capture system for novices. Instead of capturing the precise motion of professional actors, we capture the approximate motion of abstract physical proxies that users manipulate to \emph{convey} the desired animation.
We let users manipulate arbitrary everyday objects, \rev{covering rigid, articulated, and deformable items, e.g., pens, fruits, humanoid puppets, dolls}. Then we capture their performance with a single web camera, from which we locate proxy parts and track their 3D motion using modern computer vision models. 
However, not only is such a low-cost capturing system imprecise, but manipulating everyday objects inherently provides only a simplified depiction of character animations. As a consequence, directly applying the captured motion to virtual characters would result in uncanny animations that lack many of the detailed, secondary motions that make character animations realistic. To address this challenge, we complement the user performance with realistic motion priors learned from a large dataset of high-quality human motion capture. Specifically, we treat the proxy motion as an input condition to a generative motion model, which produces a character animation that follows the approximate proxy animation while augmenting it with realistic secondary motion.
\rev{Table~\ref{table:compare1} provides a comprehensive comparison of \methodName with the existing motion authoring tools from physical input. Our method is the first capable of generating \emph{high-quality motion }from \emph{any object} with merely \emph{one RGB camera}. Comparing against previous data-driven approaches, the generative model we build upon has been trained on a magnitude larger dataset and generalizes to unseen input without test-time training (please see Sec. F in the supplementary for details)}.
\begin{table*}[h]

\caption{\rev{Comparison against existing physical motion authoring systems.}}
\label{table:compare1}
\begin{tabular}{l|l|l|l|l|l}
\toprule
\rowcolor{llightgray}
\textbf{Method}  & \textbf{Capture device}               & \textbf{Physical Input}            & \textbf{Motion quality}   & \textbf{Virtual Outpu} & \textbf{Data-driven} \\ \hline
\begin{tabular}[c]{@{}l@{}}  \revv{KinEtre}~\cite{chen2012kinetre}\end{tabular} & Depth camera               & Human body                     & low               & Objects       & No          \\ \hline
\rowcolor{llightgray}
\revv{3D puppetry}~\cite{held20123dkinect}                                                                                                                              & Kinect                     & Rigid objects              & low               & Objects          & No          \\ \hline
\begin{tabular}[c]{@{}l@{}} \revv{Puppeteer}~\cite{hung2022puppeteer}\end{tabular}                                & RGB camera                 & Human body                      & low               & Human          & Yes         \\ \hline
\rowcolor{llightgray}
\begin{tabular}[c]{@{}l@{}} \revv{Motionmontage}~\cite{gupta2014motionmontage}\end{tabular}                                                            & Kinect                     & Rigid objects             & low               & Objects     & No          \\ \hline
\revv{Fingerpuppet}~\cite{hung2024fingerpuppet}                                                                                                           & RGB camera                 & Hand                       & low               & Human           & Yes         \\ \hline
\rowcolor{llightgray}
\begin{tabular}[c]{@{}l@{}} \revv{AniCraft}~\cite{li2024anicraft}\end{tabular}                            & RGB cameras                & Markers + skeleton & low                & Animals                & No          \\ \hline
\revv{Numaguchi et at.}~\cite{Numaguchi2011}    & Specific tangible hardware & Specific tangible hardware & high                   & Human          & Yes         \\ \hline
\rowcolor{llightgray}
\begin{tabular}[c]{@{}l@{}}\revv{Tangible Avatar} \cite{saint2023tangible}\end{tabular}                         & Specific tangible hardware & Specific tangible hardware & low              & Human          & No          \\ \hline
\begin{tabular}[c]{@{}l@{}}\revv{ARAnimator}~\cite{ye2020aranimator}\end{tabular}                                                    & Phone                      & Phone                      & low              & Human           &   Yes          \\ \hline

\rowcolor{llightgray}
\begin{tabular}[c]{@{}l@{}} \revv{Coolmoves} \cite{ahuja2021coolmoves}\end{tabular}                                                    & VR Controller                      & VR Controller                      & high              & Human           &   Yes          \\ \hline

Ours    & RGB camera                 & Any object           & high                  & Human          & Yes         \\ \bottomrule
\end{tabular}

\end{table*}

\begin{figure}[!t]
    \centering
    \includegraphics[width=\linewidth]{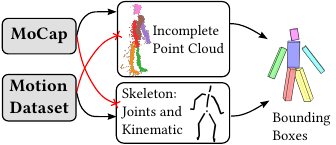}
    \caption{Our lightweight motion capture module produces noisy, partial point clouds, which we need to relate to the virtual skeletons used to represent character animation datasets. Extracting a clean skeleton from a point cloud, or synthesizing a defect-laden point cloud from a skeleton, are two difficult tasks. Our key observation is that abstract bounding boxes form a suitable middle-ground representation as they are easy to extract from both point cloud and skeleton data.}
    
    \label{fig:bbox}
    \Description{A figure clarifying the design choice of bounding boxes.}
\end{figure}

While conditional generative models have been successfully employed for related problems, such as sketch-based image synthesis \cite{zhang2023adding}, sketch-based 3D modeling \cite{SketchDream,zheng2023locally}, and sketch-based character animation \cite{zhong2025sketch2anim}, applying this technology in our context raises specific challenges. Training a conditional generative motion model requires a large dataset of paired conditions and corresponding animations, which in our case would be very difficult to acquire since creating a proxy animation and its corresponding realistic animation implies significant human labor. 
Our goal of supporting arbitrary objects as proxies further complicates this task, as the number of possible combinations of animations and proxies is virtually infinite. Our solution to this challenge is to synthesize training data by converting a dataset of realistic character animations into a dataset of proxy animations, and to bridge the gap between our synthetic data and the captured proxy motion by expressing both with the same, abstract representation.
Specifically, we represent proxy motions with \emph{bounding boxes} of object parts (see Fig.~\ref{fig:bbox} for a visual illustration). 
We describe how to extract such boxes from the input video and from the training motion capture data, and how to condition a motion generative model on proxies represented by a varying number of boxes.

We evaluate \methodName through a user study consisting of two tasks: a replication task with provided puppets and a creative task using unexpected objects proposed by participants. The results show that, with simple and intuitive user interaction (\ie manipulating everyday objects), our system can effectively generate realistic character motions that align with user intentions. Furthermore, the study also reveals users’ preferences toward different puppets, providing design implications for future input devices.
We also demonstrate how our motion capture system can be extended to keyframe-based animation for finer-control on character poses.

To conclude, this paper contributes to puppetry-guided animation creation in several aspects:
\begin{itemize}
    \item We introduce a lightweight, vision-based system for puppetry digitization, allowing the motion capture of arbitrary objects with a single web camera. %
    \item We describe how to complement approximate proxy animations with data-driven human motion priors to produce realistic character animations through puppetry.
    \item Our user study provides insights into system effectiveness, puppet preferences, and further features in demand for puppetry-based animation creation.
\end{itemize}

\section{Related Work}
We focus our discussion on prior research in using physical proxies to control character animations. We also introduce recent work in computer vision and generative models that enables our lightweight motion capture system.

\subsection{Character Animation with Physical Proxies}
The difficulty of using professional animation software is in part due to the challenge of manipulating individual body parts in a 2D interface to specify the 3D pose of human characters. This difficulty has motivated research about using various forms of proxies to define character poses directly in the physical world.

\rev{\textbf{Dedicated tangible devices with physical sensors.} }Inspired by the ubiquity and playfulness of puppets, several works proposed custom tangible devices composed of limbs, joints and sensors that assemble to represent various skeletons, offering a direct mapping to the virtual skeleton to animate~\cite{glauser2016rig, lamberti2017virtual, jacobson2014tangible, yoshizaki2011actuated,Numaguchi2011,Dinosaur1995,saint2023tangible, ahuja2021coolmoves}.
Our work follows a similar motivation, but we aim at letting users manipulate everyday objects rather than a specific device. In particular, we leverage computer vision algorithms to track object parts instead of relying on numerous sensors placed over the object being manipulated. Moreover, while many previous systems assume that the input device has the same topology as the skeleton to be animated, we designed our system to support proxies with a varying number of moving parts, and we rely on data-driven priors to turn such partial input into realistic human motions.

\textbf{Human body parts.} Human bodies can perform humanoid character motion naturally. Motion capture systems~\cite{desmarais2021review, moeslund2006survey, shimada2020physcap, wang2024egocentric} have been widely studied to extract humanoid skeleton motion from actor performances. These systems typically involve rather complex environment setups, e.g., calibrated cameras, markers, and green screens. Another branch of work~\cite{chen2012kinetre, rhodin2015generalizing, jiang2023handavatar, hung2024fingerpuppet, hung2022puppeteer, lockwood2012fingerwalking} treats human body parts as proxies to control non-humanoid virtual characters. For example, \citet{chen2012kinetre} reconstruct the human body and link it to various digital characters, controlling character motions through body movements. \citet{rhodin2015generalizing} track body, facial, and hand motions of a human to control non-humanoid characters. \citet{jiang2023handavatar} optimize hand-to-avatar joint-to-joint mappings based on user-defined calibration for real-time control of characters. 
\rev{ FingerWalking~\cite{lockwood2012fingerwalking} and FingerPuppet~\cite{hung2024fingerpuppet} similarly show that low-DoF hand performances can act as intuitive proxies for producing full-body locomotion. }
 Though the input of human body actions is accessible, mapping human body parts to non-humanoid characters inherently produces unnatural motions, since the topology and kinematics between humans and those characters are different. While we demonstrate our approach on human characters, we face a similar challenge as this family of work as we seek to map the abstract motion of simple proxies to the realistic motion of complex skeletons. \rev{While optimization-based techniques have been studied to synthesize physically-valid motion~\cite{fang2003efficient}, \methodName instead leverages data-driven motion priors to generate realistic and expressive animations.
}

\rev{\textbf{Character-like daily objects.}} Several studies~\cite{li2024anicraft, ye2020aranimator, gupta2014motionmontage, held20123dkinect, sin2022tracking} offer solutions to produce virtual animations by manipulating common objects. 3D Puppetry~\cite{held20123dkinect} reconstructs rigid objects with a depth camera and maps manipulation of the object to the rigid motion of its digital twin. 
\rev{Similarly, P. T. Sin et al.~\cite{sin2022tracking} track a non-rigid stuffed toy to control a virtual model of the same toy in XR games.}
ARanimator~\cite{ye2020aranimator} uses a cell phone as a 6 DoF sensor and builds a regression model based on a user study to predict motion types from cellphone movements. %
AniCraft~\cite{li2024anicraft} proposes an affordable proxy for prototyping 3D character animation in MR. The process involves handcrafting the desired skeleton with metal wires and markers. Then, the manipulation of the proxy is mapped to skeleton motion automatically. 
While lightweight and accessible, these methods share two drawbacks. First, they rely on specific objects or devices to capture the user performance. 
Second, they only support a limited space of motion, such as rigid~\cite{held20123dkinect, gupta2014motionmontage} \rev{and cage-based~\cite{sin2022tracking}} transformations, pre-defined motion templates~\cite{ye2020aranimator}, and prototyping-level unnatural motion~\cite{li2024anicraft}.
In contrast, \methodName allows users to produce realistic animations simply by manipulating common objects in front of a web camera.

\

\subsection{Vision Foundation Models}
Vision foundation models~\cite{kirillov2023segment,karaev2024cotracker3,wang2025pi,awais2025foundation} refer to a class of deep neural networks trained on massive datasets to address general vision problems, such as image segmentation, video tracking, and visual geometry reconstruction. Segment Anything~\cite{kirillov2023segment} segments an image into meaningful semantic regions based on a text prompt or user clicks, while CoTracker~\cite{karaev2024cotracker3} takes videos as input and tracks pixel correspondences across all frames. 
More recently, VGGT~\cite{wang2025vggt} and $\pi^3$~\cite{wang2025pi} have been proposed to predict 3D point clouds and camera parameters directly from images. By combining these tools, we build a motion capture system with minimal hardware requirements—only a webcam, without the need for calibration.

\subsection{Motion Generative Models}
Generative models learn to estimate real-world data distributions, and generate novel realistic data by sampling from the estimated distributions. Popular architectures include Generative Adversarial Networks~\cite{goodfellow2014generative} and Variational Auto Encoders~\cite{kingma2013auto}, with diffusion~\cite{dhariwal2021diffusion, croitoru2023diffusion} and flow-based models~\cite{lipman2022flow} showing dominating performance recently. 

Motion generative models have been trained on large motion capture datasets to generate different types of skeleton motions from different controlling signals. A major topic is text-conditioned human motion generation~\cite{tevet2022human, Guo_2022_CVPR,zhang2024motiondiffuse,azadi2023make,guo2024momask}. For example, MDM~\cite{tevet2022human} is a diffusion model based on HumanML3D~\cite{Guo_2022_CVPR} and the Kit dataset~\cite{Plappert2016}, and is guided with text conditions through classifier-free guidance~\cite{ho2022classifier}. While human motion is the dominant problem of interest due to data availability, some recent research~\cite{yang2024omnimotiongpt, gat2025anytop, song2025puppeteer} focus on generating arbitrary skeleton motion. Also, different control signals like sketch~\cite{zhong2025sketch2anim, peng2023dualmotion}, audio~\cite{xu2025mospa}, video~\cite{li2025genmo}, and joint positions~\cite{xie2023omnicontrol} have been explored. Moreover, since motion generative models posses motion priors learned from motion capturing data, recent research~\cite{wang2024egocentric} shows that such diffusion priors can be used backward to refine the motion capturing system's result.

We demonstrate our method with human motion generation models, enabling intuitive puppetry for humanoid characters' motion. Our motion capture system is agnostic to motion type, and our box-conditioned generative model could be generalized to arbitrary skeletons and character types if applied to the corresponding base generative models.

\section{Method}
\label{sec:method}

Fig.~\ref{fig:teaser}(b) displays a high-level illustration of our method with two core modules - vision-based motion capture (MoCap) and conditional motion generation (MoGen). 
Given the recorded puppetry performance video, our goal is to recover an approximate 3D animation from the video and translate it into a realistic character animation.
Specifically, we analyze the video frames to obtain the puppet parts and track the part movement to obtain the approximate motion, which, subsequently, serves as the spatial condition in the motion generation module with realistic motion priors pre-learned from large-scale motion datasets.

Due to the lack of a direct puppetry-motion dataset, there exists a mismatch between the approximate motion from the video input (\ie partial point clouds) and the realistic motion from motion datasets (\ie clean SMPL human motion \cite{Guo_2022_CVPR}).
We propose to use 3D bounding boxes as an intermediate and abstract representation to bridge the two modalities. %
\rev{The choice of bounding boxes over other primitives, such as cylinders or line segments, is motivated by their ability to represent rotations around the object’s symmetry axes. Ellipsoids could also be used, as they have the same degrees of freedom as cuboids, but we opt for bounding boxes since they are the de facto standard in computer vision tasks such as 3D object detection~\cite{mao20233d}.}
In the following, we elaborate on technical details.

\subsection{System Setup and User Input}
\label{subsec:system_workflow}
Fig.~\ref{fig:teaser}(a) shows our system setup, where a web camera (\eg a mobile phone) is mounted on a desk (as the ground plane), facing the performing space. 
A checkerboard marker is placed on the ground plane for calibrating the up direction of the ground.
Users manipulate a physical proxy, such as a humanoid puppet, a toy, or even their fingers, to perform the intended motion.
After the performance, the recorded video serves as the input to our algorithm\rev{, along with an optional text description of the desired motion}.
Note that we tested our system in an indoor environment with artificial lights or natural daylight. 

In the current implementation, an extra user input is necessary. Given the first frame of the recorded video, users are asked to click a point on the ground object, and a few points on the different parts of the proxy  \rev{they wish to articulate}. \rev{These points are fed to a video segmentation model, as explained in the next section.}
Note that users do not need to segment the proxy into many small parts: \rev{1-6} parts are often enough to convey the motion of the legs, torso and arms of a character. \rev{In supplemental materials, we illustrate a typical interactive segmentation session and visualize the user clicks for all results shown in the paper.}

\begin{figure*}[!tb]
  \includegraphics[width=\textwidth]{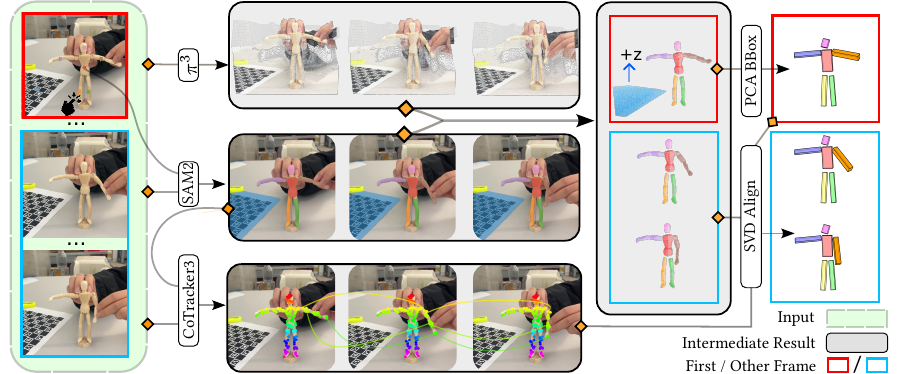}
  \caption{
  Overview of MoCap. From left to right:  {given the recorded video and user clicks in the first frame, we exploit SAM2~\cite{ravi2024sam} to segment the parts of the puppet in all frames. \rev{ The user clicks indicate the desired part segmentation, with different colors denoting different parts. The order of the clicks does not influence the result. We also run }
  $\pi^3$~\cite{wang2025pi} to estimate the point cloud of each frame, and CoTracker3~\cite{karaev2024cotracker3} to produce dense pixel-wise correspondences between frames.
  Combining semantic segments, point clouds, and motion tracks allows us to recover 
  3D bounding boxes of proxy parts and their motion across the video clip.} 
  }
  \label{fig:pipelineMocap}
    \Description{A diagram showing the technical steps of the first motion capture module in our method.}
\end{figure*}

\subsection{Motion Capturing with Vision Foundation Models}
\label{subsec:mocap}

The goal of our motion capture module is to \rev{segment} proxy parts, reconstruct their 3D pose in each video frame, and track the movement of the parts across frames. Fig.~\ref{fig:pipelineMocap} shows an example video, illustrating the main technical steps of our approach.

\textbf{3D reconstruction with $\pi^3$~\cite{wang2025pi}.}
To reconstruct the 3D information of the dynamic scene, we exploit $\pi^3$, running on each video frame to produce a 3D point cloud of each frame (Fig.~\ref{fig:pipelineMocap}, row one). 
Particularly, the point cloud takes the form of a depth map, providing point-pixel alignment, which facilitates the following segmentation and tracking propagation.

\textbf{Segmentation with SAM2~\cite{ravi2024sam}.}
\rev{We exploit SAM2~\cite{ravi2024sam} to segment proxy parts over the entire video. This segmentation model takes as input the user-provided points in the first frame to extract the corresponding segments over all frames of the video (Fig.~\ref{fig:pipelineMocap}, second row). The model is robust to occlusions, producing a consistent segmentation even when parts of the proxy are not visible in some of the frames.}
Additionally, we leverage the pixel-point correspondence from $\pi^3$ to transfer the part segments onto the point cloud of each frame, effectively producing segmented point clouds.
We rely on the segmented ground object (\ie the checkerboard) to rotate the global z-axis from $\pi^3$ so that it aligns with the upward ground direction.

\textbf{Tracking with CoTracker3~\cite{karaev2024cotracker3}.}
In our formulation, we assume the motion to be rigid, since most everyday objects can be reasonably approximated as articulated rigid bodies.
With the segmented parts in the per-frame point clouds, there are several potential solutions for inferring the motion information. 
For example, point cloud registration techniques can be used to estimate the per-part rigid transformation. However, there are two major difficulties. Firstly, the point cloud of each part is very sparse, as each part occupies only a small region, which poses challenges for registration algorithms. Secondly, these algorithms are highly sensitive to point noise introduced by imperfect segmentation and reconstruction near part boundaries.

Our solution is to resort to tracking pixels on video frames for a robust motion estimation.
Specifically, we run CoTracker3~\cite{karaev2024cotracker3} on the input video. \rev{This tracking model takes as input so-called \emph{query pixels} in a given frame of the video and tracks them across all frames. In our system, we automatically sample query pixels within each segment} to produce tracks throughout the video (see Fig.~\ref{fig:pipelineMocap}, third row). \rev{In most cases, sampling these pixels in the first frame suffices to track object parts over the entire video. But the tracks might get lost in the presence of occlusions. Fortunately, the segments extracted by SAM2 are robust to occlusions, which allows us to resample the segments in subsequent frames where they appear to handle these more complex cases. Specifically, the resampling is automatically triggered repetitively when 70\% of tracking pixels becomes invisible (see Fig. 4 in the supplementary for an example).}
Thanks to the pixel-point correspondence from $\pi^3$, we transfer the tracks from 2D pixels to 3D points. This process gives us, for each segmented part, a one-to-one point mapping between successive frames over a subset of the part points.

\textbf{Motion with bounding boxes.}
Having the segmented point clouds with robust tracks in each of them, we next estimate the motion with bounding boxes.
For the initial frame, we first filter out the low-confidence points and outliers based on their confidence scores predicted by $\pi^3$. We then compute an oriented bounding box through PCA estimation (see top-right of Fig.~\ref{fig:pipelineMocap}). 
Starting from these boxes, we rely on established point tracks to estimate the transformation through SVD alignment (Kabsch-Umeyama algorithm~\cite{umeyama2002least}), and transform the initial boxes to later frames, forming the motion sequence represented by bounding boxes (see the last column of Fig.~\ref{fig:pipelineMocap}).
\rev{Note however that the system cannot estimate the transformations of the bounding boxes in the frames where the corresponding segments are fully occluded. We handle these cases by randomly removing some of the boxes during training of the conditional motion generation model, such that the model learns to synthesize plausible motion for the missing parts. Please see the supplementary for a dedicated analysis of the robustness of our capture system under occlusion and in-place rotation.}

\begin{figure*}[!tb]
  \includegraphics[width=\textwidth]{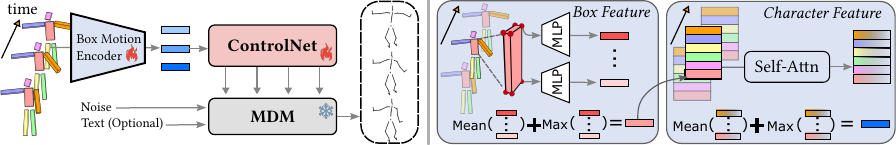}
  \caption{
   {Left: overview of our conditional motion generator. We train a custom box motion encoder alongside a ControlNet module to condition a pre-trained Motion Diffusion Model. Right: for a given frame, our box motion encoder first encodes each vertex of a box (\eg the pink one) using an MLP, and aggregates all vertices of the box into a single code using mean and max operations to obtain a latent code that is invariant to vertex ordering. A self-attention layer then exchanges information between the latent codes of all boxes of the character proxy. Finally, the resulting latent codes are aggregated into a single code for the entire character, again using mean and max to be invariant to box ordering.}
  }
  \label{fig:pipelineMoGen}
  \Description{A diagram showing the overview of the motion generation module and the illustration of our novel box motion encoder.}
\end{figure*}

\subsection{Box-guided Motion Generation}
\label{sec:moGen}

Conditioned on the coarse motion extracted from the performance video, we aim to generate the corresponding realistic character motion exploiting the motion priors in learned models. \rev{Given the segmented 3D point cloud, a straightforward solution would be to convert every of its part into skeletal joints. However, this solution is only feasible when the number of parts corresponds to the number joints of a standard skeleton, which is rarely the case in our setting where everyday objects contain only a few moving parts. Furthermore, directly reconstructing a skeleton animation from the captured point clouds would yield uncanny results, as it would strictly replicate the user's imperfect physical manipulation and lack the realistic motion effects provided by generative priors. We thus introduce bounding boxes as an intermediate representation to bridge the input 3D points with realistic output motion.} In the current implementation, we only target human-like motion due to the wide availability of existing human motion datasets.

\textbf{Preliminary.}
 {Motion diffusion models 
typically represent human motions with a fixed number of body part joints.}
In addition to text conditions, spatial guidance such as 3D joint trajectories or keyposes can be injected into the motion diffusion process via the ControlNet~\cite{zhang2023adding} mechanism to further control the behavior of the generated motion \cite{xie2023omnicontrol,zhong2025sketch2anim}. 
 {Additionally, the accuracy of the generated motion with respect to the spatial conditions can be enhanced by a so-called inference guidance loss~\cite{xie2023omnicontrol,zhong2025sketch2anim}, inspired by classifier guidance~\cite{dhariwal2021diffusion} and loss-guided diffusion~\cite{song2023loss,wang2023intercontrol}}. 

\textbf{Method overview.} Built upon a pre-trained human motion diffusion model - MDM~\cite{tevet2022human}, we design our method as a conditional motion generation, as shown in Fig.~\ref{fig:pipelineMoGen}. 
Briefly, the generator takes as input the extracted motion guidance and an \emph{optional} text description, to denoise a Gaussian noise into a realistic motion with an inference guidance loss.
However, in our setting, there are two unique challenges due to the bounding box-based intermediate motion representation.
Specifically, different physical proxies express different levels of abstractions of character parts. For example, the banana in Fig.~\ref{fig:teaser} has four boxes representing the upper body, lower body, and two arms, while the humanoid puppet in Fig.~\ref{fig:pipelineMocap} has six boxes representing the head, body, two upper limbs, and two lower limbs. 
On the one hand, the number and the order of boxes vary depending on the proxy \rev{and the number of parts segmented by the user}. On the other hand, the mapping from boxes to body part joints is infeasible to build explicitly. This prevents a one-to-one guidance between boxes and joints and also exposes difficulty in formulating the inference guidance loss term.
To overcome these challenges, we have designed a novel permutation-invariant box motion encoder and an innovative box-joint guidance. 

\textbf{Box motion encoder.}
\rev{A key idea behind our system is to summarize the input bounding boxes -- which can vary in number depending on the number of parts segmented by the user -- into a compact fixed-size motion code.}
We design this encoding process at both the box level and the character shape level  to be invariant to vertex and box ordering. \rev{Permutation invariance is essential to extract a unique feature representation for a bounding box, as the order of vertices does not alter the underlying geometry. We employ Mean/Max pooling, which are proven effective for general point cloud geometry learning (e.g., PointNet~\cite{qi2017pointnet}). 
 The network architecture is shown in Fig.~\ref{fig:pipelineMoGen} (right).}
Firstly, for each 3D bounding box within a frame, we employ a shared MLP to encode the 3D coordinates of each vertex into a latent feature. To make the box embedding invariant to vertex order, we aggregate the features by summing up the mean and maximum feature values of the eight box vertices.
We repeat the box encoding for each box of the proxy.
Secondly, for proxies made of several boxes, 
we exploit the self-attention mechanism to model the inter-box relationship within the frame, and then employ the same aggregation strategy to produce the shape-level feature for that frame.

\begin{figure}[t]
    \centering
    \includegraphics[width=\linewidth]{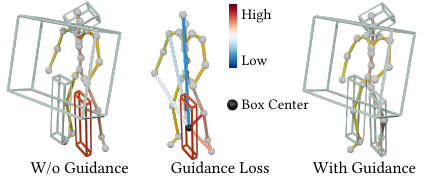}
    \caption{Impact of spatial guidance and its design illustration. 
    Left: without spatial guidance, the bounding boxes are able to provide rough motion control, but do not guarantee precise alignment between boxes and joints (\eg the joints on the right leg vs. the red box). 
    Middle: for an exemplar bounding box (in red), the guidance measurement loss is computed against every joint, then accumulated using distance-based combination weights.  {Five} such distances are visualized, with colors indicating the corresponding weights.
    Right: With the designed spatial guidance, the generated motion aligns closely with bounding boxes, ensuring each box contains at least one joint.
    }
    \label{fig:guidance}
        \Description{A figure showing the impact of with and without spatial guidance, and its design intuition.}
\end{figure}

\textbf{Box-based spatial guidance.}
Users manipulate the physical proxy to drive character motion, even when represented by a single bounding box (\eg a pen or a book). In this sense, the boxes are expected to span all the joints, despite the absence of a direct box-joint mapping. 
Thus, the core idea of defining the discrepancy measurement is to make sure that \emph{every bounding box contains at least one joint} (see Fig.~\ref{fig:guidance}, for an illustration).
Specifically, given joints from one frame of the produced motion and corresponding conditional boxes, we measure the distance between every joint to the center of every box.
Instead of setting a distance threshold to determine the coverage relationship, we use a \emph{soft} association between each box and all the joints, and the weight decays exponentially with distance.
Finally, we sum up all the weighted distances linearly and minimize the total value as the loss guidance at inference. 

\textbf{Network training and inference.}
The MDM~\cite{tevet2022human} was pre-trained on the HumanML3D~\cite{Guo_2022_CVPR} dataset. 
As shown in Fig.~\ref{fig:pipelineMoGen}, we only train the box motion encoder and the ControlNet on our constructed dataset, where each data item consists of the paired abstracted box motion and realistic skeletal motion. 
For robust network training, we adopt several data augmentation techniques. \rev{Specifically, we vary the number of boxes to achieve different levels of details}, we randomly apply rigid transformations to a few boxes in a random frame of a motion sequence  {to mimic tracking failure}, and we randomly drop either the text input or a few boxes {to simulate different levels of abstraction in the condition}.
At inference time, given the detected conditional box motion sequence, the initial noise, and the optional text, our motion generator produces the corresponding realistic motion, conforming to the spatial conditions.
\emph{Please refer to the supplementary for our dataset construction process and implementation details of our method.}

\section{User Experience Study}
\label{sec:user_study}
To evaluate how \methodName facilitates character motion creation and assess user preferences across different interaction proxies, we conducted a comprehensive user study with diverse participants and task designs, and a questionnaire-based feedback collection.

\subsection{Study Methodology}

\paragraph{Study participants}

We recruited 9 university students aged 20 to 30 years to participate in our study. The participants have mixed backgrounds and fluency with animation creation tools, with 4 having no prior experience, 3 possessing entry-level experience, and 2 familiar with a few software (\ie Blender, Cinema4D, 3ds Max and Maya).

\paragraph{Experimental setup}
The study was conducted in a controlled laboratory environment featuring a desk-mounted setup with a smartphone serving as the capture device. Participants had access to various physical puppets for motion manipulation. All captured motion video was transmitted to a remote server equipped with an RTX 4090 GPU for \emph{offline} processing and animation generation.

\paragraph{Study protocol}

The study followed a structured four-phase protocol before a post-study evaluation.
\begin{itemize}
    \item[1)] \textbf{Pre-study assessment}: Participants completed an initial questionnaire documenting their background and experience with 3D animation tools. They are also asked to estimate how hard the replication task could be without our tool.

    \item[2)] \textbf{System introduction}: We provided a 10-minute tutorial covering fundamental animation concepts and the \methodName system workflow, ensuring all participants had equivalent baseline knowledge.

    \item[3)] \textbf{Replication tasks}: Participants performed three motion replication exercises, where they recreated target animations using the provided tools and puppets. These tasks assessed the system's usability for reproducing specific motions. \revv{ A default short text description is applied in this task.}

    \item[4)] \textbf{Creative design tasks}: Following the replication tasks, participants engaged in open-ended creative animation design, selecting from available puppets or incorporating their own objects to create original character motions. This phase evaluated the system's potential for creative expression. \revv{An optional text description is provided by the user.}

    \item[5)] \textbf{Post-study evaluation}: After reviewing their generated animations, participants completed a comprehensive questionnaire and participated in a semi-structured interview to gather qualitative feedback about their experience and system performances.
\end{itemize}
\emph{All the raw data from the user study can be found in the supplemental material.}

\begin{figure*}[!tb]
    \centering
    \includegraphics[width=\linewidth]{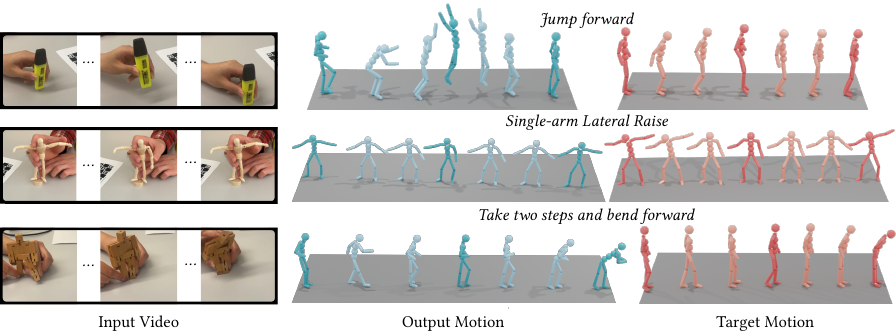}
    \caption{Representative replication results from our user study. Given a target motion (right), participants were asked to reproduce it (middle) by manipulating a designated physical proxy (left).
    }
    \label{fig:user_replication}
    \Description{A figure showing the user replication results with three typical examples. From left to right, the recorded video frames of a user manipulating a given puppet, the produced and the reference motions are presented, respectively.}
\end{figure*}

\begin{figure*}[!tb]
    \centering
    \includegraphics[width=\linewidth]{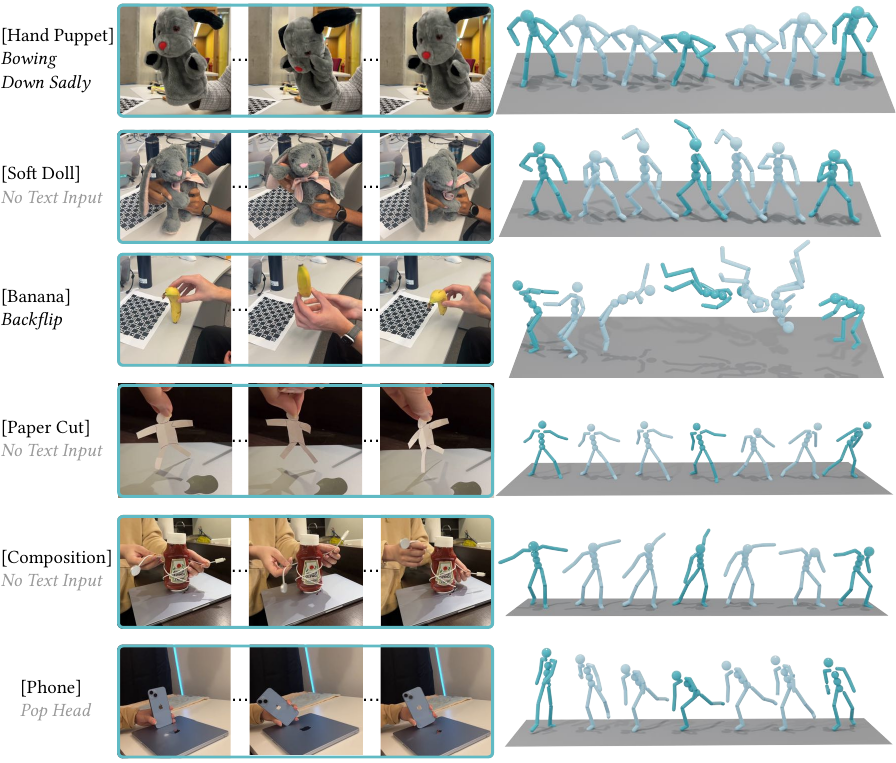}
    \caption{Creative results from participants manipulating diverse puppets (first three rows), \rev{and from an experienced user performing the action using paper cut (fourth row),  compositing daily objects (fifth row), and a phone (last row)}, demonstrate the effectiveness of our system in producing high-quality motion \rev{from varying physical proxy}. The left column shows the puppets and input texts (if any), the middle column shows the input video frames, while the right column presents the corresponding generated motions. Note that the text input is provided only in the first and third examples. For better visualization, we display the motion frames with spatial displacements on the ground, although these displacements are not part of the motion itself (\eg the motion in the second last example is basically moving in place).
    }
    \label{fig:gallery}
        \Description{A figure demonstrating diverse and compelling motions produced by our system.}
\end{figure*}

\begin{figure}[!t]
  \centering
  \includegraphics[width=\columnwidth]{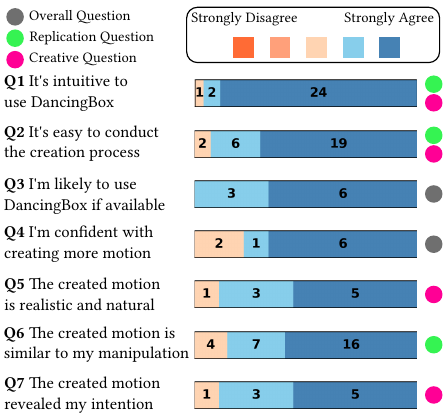}
  \caption{\textbf{Questionnaire statistics (subset).}
    A selected subset of questions with corresponding responses is shown. Gray, green, and magenta colors indicate question categories asked at different stages:
    \emph{overall} questions were asked once after the study; 
    \emph{replication} questions were asked after each replication task, yielding \(3 \times 9 = 27\) responses;
    \emph{creative} questions were asked after the creative task with \(1 \times 9 = 9\) responses.
    Some questions were asked after both the replication and creative tasks, with \(27 +9 = 36\) responses.
    Each question has five response options (\ie from Strongly Disagree to Strongly Agree, scored 1 to 5).
    Numbers on the bars indicate the count for each response (colored accordingly). 
    Notably, across all selected questions, no participant chose \emph{Disagree} and \emph{Strongly Disagree}. 
    See the supplementary for the complete set of questions and responses.
    }
  \label{fig:stastic}
   \Description{A figure showing the questionnaire statistics on a selected subset of user study questions. For each question, five scores from Strongly Disagree to Strongly Agree can be chosen.}
\end{figure}

\subsection{System Usability and Result Motion Quality}
\label{subsec:user_results}

Figs.~\ref{fig:user_replication} and \ref{fig:gallery} present representative results from participants for both replication and creative tasks. 
Overall, all participants successfully reproduced the given motions.
Notably, when asked \emph{before} the study to estimate how long it would take to create motions similar to the three examples with existing tools (Fig.~\ref{fig:user_replication}), even the two most experienced participants anticipated needing at least \emph{one full day}. 
In contrast, with our system, the operating time to reproduce a single motion is  {roughly 3 minutes for puppet manipulation, 2--4 minutes for interactive part clicking, and 4--5 minutes for the algorithm running (see Sec.~\ref{sec:results} for a detailed breakdown).}
In the creative design stage, participants explored a variety of puppets to produce diverse motions, including unexpected outcomes such as a hand puppet crying, a soft-bodied doll dancing, and a banana performing a backflip (the first three examples in Fig.~\ref{fig:gallery}, respectively). We summarize the findings below.

\paragraph{Users found the system extremely easy to use.}
Question responses indicate that participants found the system intuitive to learn (Q1, Fig.~\ref{fig:stastic}) and easy to operate using physical proxies (Q2, Fig.~\ref{fig:stastic}). As P3 put it, ``\textit{I used to play with dolls as a kid---this felt just as easy.}'' All participants reported that they would use the tool for animation creation if available (Q3, Fig.~\ref{fig:stastic}). P7, who was working on a personal animation project, remarked: ``\textit{I wish your system were ready right now; it's much easier than my usual software (Blender).}'' After completing both replication and creative design tasks, all participants believed they could create additional motions with the system (Q4, Fig.~\ref{fig:stastic}).

\paragraph{Users reported natural, realistic motions.}
The fifth question (Q5, Fig.~\ref{fig:stastic}) shows that participants perceived the generated motions as realistic and natural. Leveraging a generative prior learned from rich motion data, \textit{DancingBox} can produce realistic motion even when proxy movements are somewhat unnatural. P8 commented: ``\textit{I thought the rabbit's head (second example in Fig.~\ref{fig:gallery} ) wasn't moving the right way, but the final motion turned out perfect.}''

\paragraph{The system respects user intent.}
A common issue with generative methods is unintended drift from the user's intent. We evaluated how well the system preserves user intent across both tasks. In the replication task, participants rated the similarity between their manipulations and the generated motion with an average agreement score \emph{4.44} (Q6, Fig.~\ref{fig:stastic}), indicating strong faithfulness to the proxy. In the creative task, participants produced various motions with their chosen tools, and most reported that the results reflected their intentions (Q7, Fig.~\ref{fig:stastic}). P3, P5, P4, and P8 each expressed a similar sentiment: ``\textit{This is exactly what I had in mind.}''

\paragraph{Our system facilitates motion replication.}
Imitating existing motion from video or image references (whether designed or recorded) is a common practice for creating skeleton motion in both motion capture pipelines and 3D software. We quantified how well our system fits this replication workflow in the user study. During the replication task, participants rated pairwise similarity among the target motion, their manipulation, and the generated motion. To evaluate effectiveness, we define a \emph{replication failure} as any case where the generated motion is \emph{less} similar to the target than the original manipulation:
\[
\mathrm{Sim}(\text{generated}, \text{target}) < \mathrm{Sim}(\text{manipulation}, \text{target}).
\]
Across 27 trials (9 participants $\times$ 3 replication tasks), we observed only 2 failures (7.4\%), indicating that in over 92\% of cases the system preserved or improved similarity to the target motion.
These results suggest that, in replication scenarios, when proxies approximate the target motion reasonably well, our system reliably translates them into detailed motion sequences while maintaining similarity.

\subsection{Implications for Physical Proxy}

Our system supports diverse puppets---from simple rigid objects to soft dolls with virtually unlimited degrees of freedom. We further examined how participants' perceptions of different puppet types, providing insights for the design of future puppets and tangible interfaces.

\paragraph{Users prefer detailed puppets and want fine control.}
In the questionnaire, when asked to choose between abstract puppets (e.g., a pen, a banana, or a hand pose) and a detailed articulated human model, eight out of nine participants chose the articulated model. But the reasons varied. First, detailed puppets directly resemble humans, reducing cognitive overhead; as P2 noted, ``\textit{They’re similar to real characters and give immediate feedback about the motion.}'' Second, human-like puppets afford fine-grained control. P8 said, ``\textit{I can’t really show hand movements with a pen, so I’d rather use a doll here.}'' P7 added, ``\textit{Abstract puppets can work for stage play, but I want to control every single detail with the human model.}''

\paragraph{Controlling detailed puppets can be hard.}
P9—the only participant who preferred abstract puppets—explained, ``\textit{The pen and the banana were the easiest, mainly because it's hard to manipulate the humanoid puppets accurately in a short time}.'' Other participants, while favoring detailed control, acknowledged that detailed puppets can be challenging. As P7 put it, ``\textit{That wooden tool is harder to use. $\cdots$ I'll need a third hand.}'' When asked about the sources of difficulty, participants mentioned ``\textit{the joints are lagging}'' (P5), ``\textit{the puppet easily falls over, so I have to keep it upright while moving}'' (P3), and ``\textit{having to manage occlusion during manipulation}'' (P2).

To conclude, participants generally prefer detailed, human-like puppets for expressive control, yet recognize they can be challenging to operate; abstract puppets remain useful for scenarios like stage-play animation. Because our \methodName system is agnostic to puppet type, it can support future exploration of input devices (puppets and other physical proxies as tangible controls). Our findings suggest that a balance should be struck between finer detail and controllability: adding too many flexible joints may reduce practical usability if users cannot physically control all of them.

\section{Results and Discussions}
\label{sec:results}

Beyond the user study, we also demonstrate more appealing motions results created by an experienced user, as shown in Figs.~\ref{fig:teaser} and \ref{fig:gallery}. 
In this mode, our system effectively functions as a markerless motion capture pipeline: it takes an actor’s performance video as input and outputs a skeleton motion sequence. \rev{The system is calibrated by placing a planar object to denote the ground, followed by a few clicks during segmentation.}. Compared with established motion capture systems, our system offers two advantages: 1) it requires only monocular video from a consumer webcam, with no markers or suits; and 2) it can be used immediately with a smartphone and a user's own performance, with virtually no setup overhead.
\rev{Compared with previous physical animation authoring tools, \methodName can reproduce their results without any parameter tuning (Fig.~\ref{fig:teaser}(b) vs. FingerPuppet~\cite{hung2024fingerpuppet} and Fig.~\ref{fig:gallery} last row vs. ARAnimator~\cite{ye2020aranimator}). Moreover, \methodName can accommodate a much wider range of input objects, giving users substantially greater freedom for exploration.}
However, the main limitations are reduced motion accuracy compared to multi-camera systems and non-real-time processing.

To further validate the effectiveness of our method, we have conducted 
an ablation study and discussions of key design choices (Sec.~\ref{subsec:discussion}), and demonstrated an application by extending our system to support keyframe-based motion capture (Sec.~\ref{subsec:application}).
Please refer to the \emph{supplemental video} for better dynamic motion visualization.

\paragraph{Runtime efficiency.}
Excluding user interaction, the current implementation requires a non-trivial execution time on a single NVIDIA 4090 GPU. 
The complete pipeline runs in approximately 2min 40s, with the following breakdown: 
$\pi^3$ – 30s, SAM2 – 30s, CoTracker – 10s, bounding box estimation – 30s, and MoGen with spatial guidance – 60s. 
In addition, hardware constraints (\eg GPU memory) introduce further overhead (about 2 minutes), like loading and unloading vision models 
(see the discussions in Sec.~\ref{subsec:discussion}).

\subsection{Ablation Study and Discussions}
\label{subsec:discussion}

\begin{figure}[!t]
    \centering
    \includegraphics[width=\linewidth]{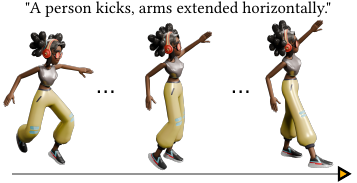}
    \caption{Generation result from the same video input of Fig.~\ref{fig:teaser}(b), but adding extra text description to control the arms. 
    }
    \label{fig:textControl}
    \Description{A figure showing a generated motion with an extra text as input. The text control complements the box guidance.}
\end{figure}

\paragraph{Impact of text input.} 
In our MoGen, the impact of the optional input text varies depending on the abstraction level of the bounding box:
\begin{itemize}
    \item High abstraction level: as shown in the third example in Fig.~\ref{fig:gallery}, a single box covering the whole banana. In this case, the text input is \emph{indispensable}. It helps constrain the generation space; otherwise, from a single rotating box, it is impossible to reason out the motion of a ``Backflip''. 
    
    \item Medium abstraction level: in Fig.~\ref{fig:teaser}(b), the character's lower limbs are explained via several boxes, but the whole upper body, including two arms, is described with a single box, without the motion control. In this case, the input text is \emph{complementary} to the box control by specifying the action of two arms. See the generated motion in Fig.~\ref{fig:textControl} with two arms lifting horizontally.
    
    \item Low abstraction level: the humanoid puppet in the second example of Fig.~\ref{fig:user_replication} has all the body parts covered by a separate box. In this case, the text input is \emph{impactless} as the boxes explain the motion well.

\end{itemize}

\paragraph{Occlusion.}
Because the input is monocular, our MoCap module is sensitive to occlusion. Occluded regions may be masked out by SAM, yielding incomplete point clouds. Moreover, CoTracker can lose track when large occlusions occur during puppetry. Since puppetry necessarily involves hands, complete avoidance of occlusion is difficult. A natural mitigation is multi-view capture, adding cameras at different angles. Generative completion is another direction: remove (or inpaint through) the occluding hand while reconstructing the hidden puppet geometry and motion. Participants also suggested practical aids such as ``\textit{automatically warn the user when important regions are occluded}'' and ``\textit{use string puppets so hands are not in frame}''. We leave these avenues to future work.

\paragraph{Toward real-time use.} 
Although the current system runs  {non-real-time}, several components can be substantially accelerated:
\begin{itemize}
    \item[(a)] \emph{MoCap}. At present, three vision foundation models run sequentially to fit within a single NVIDIA 4090 GPU, and CoTracker ingests point samples taken from SAM masks to save memory. A significant fraction of runtime is spent on (i) \textit{model swapping}—repeatedly loading/unloading large model weights between CPU and GPU, and (ii) \textit{staging}— saving intermediate outputs (masks, point tracks, point clouds) to disk, and disk{-}CPU{-}GPU transfers to hand results from one module to the next. With sufficient GPU memory (\(\sim\)100~GiB), the three models can run at the same time without switching and can share data directly on the GPU. 
    In that case, CoTracker can track a uniformly sampled set of points directly, followed by trajectory filtering using SAM masks.

    \item[(b)] \emph{MoGen}. We currently use vanilla MDM with 1000 diffusion steps for sampling. Following established step-distillation approaches~\cite{dai2024motionlcm}, this can be reduced to \(\approx20\) steps, potentially yielding \(\sim\)50\(\times\) speedups without materially affecting quality.
\end{itemize}
As the system is already complete in its current form, we defer these engineering optimizations to future work.

\begin{figure}[!t]
    \centering
    \includegraphics[width=\linewidth]{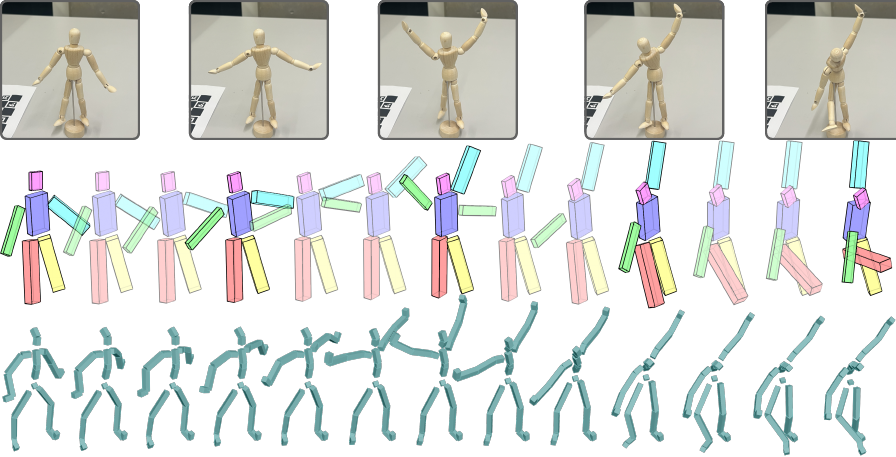}
    \caption{Top row: users specify five keyframes, indicating the motions of both arms, the torso, and one leg.
    Middle row: bounding boxes are estimated for the five keyframes using the same motion capture pipeline. Additional boxes are interpolated between keyframes and shown with transparent coloring.
    Bottom row: the complete bounding-box sequence serves as guidance to generate the final realistic character motion.
    }
    \label{fig:keyframe}
     \Description{A figure showing the process of key-framing application. Top row: 3 frames of a humanoid puppet. Middle row: estimated bounding boxes and interpolation. Down row: generated motion between key frames.}
\end{figure}

\subsection{Application}
\label{subsec:application}

Our system can be extended to support keyframe-based character animation. 
As illustrated in Fig.~\ref{fig:keyframe} (top row), the user first defines a sequence of key poses for the puppet and captures them in temporal order. 
These captured keyframes are treated as consecutive frames of a video sequence, which are then processed by the MoCap system to extract bounding boxes (middle row of Fig.~\ref{fig:keyframe}). 
To synthesize intermediate frames, we interpolate the bounding boxes of adjacent keyframes along geodesics in $\mathrm{SE}(3)$, following Park et al.~\cite{park1997smooth}. 
The playback speed can be controlled linearly by specifying the number of frames to be interpolated. 
Finally, the complete sequence of bounding boxes is provided to MoGen, which generates the final motion (bottom row of Fig.~\ref{fig:keyframe}). 

\section{Conclusion and Future Work}

We introduced \methodName, a lightweight puppetry-based motion capture system that enables novices to animate characters using everyday objects and a single webcam. By conditioning generative motion models on bounding box representation, our approach transforms coarse object manipulations into realistic character motions. A user study validates that the system is effective and intuitive, supports diverse proxies, and offers design insights for tangible interfaces.

\textbf{Future Work.}
There are inspiring directions to explore in the future.
\begin{itemize}
    \item[1)] Interaction between characters.
    The current system does not yet support direct character--character interaction. Nevertheless, such interactions can be composed by applying \methodName to each puppet independently and then merging the results using 3D spatial correlations estimated from point clouds. For example, trajectories can be temporally aligned and spatially arranged based on pairwise distances, and contact cues can be derived from the reconstructed proximity. Another complementary solution is to upgrade the motion-generation module to a multi-character model \cite{MultiPersonGen25}, enabling multiple character interaction with constraints.

    \item[2)] Generalization to non-human characters.
    Our MoGen is built upon a pre-trained motion diffusion model on human motion datasets, which limits our system to human-like motions.
    While it already covers diverse human actions, it could be further extended to non-human motions by leveraging a general-purpose motion generator \cite{wang2025x-mogen}.

    \item[3)] Feature requests from participants.
    Participants proposed several practical features to enhance the system:
    \begin{itemize}
        \item Connect the system to game engines to control in-game characters,
        \item Provide explicit speed control over the generated motion,
        \item Enable editing of specific joints at specific timestamps,
        \item Add post-processing tools to compose and blend multiple actions.
    \end{itemize}
    They also suggested a two-stage, coarse-to-fine workflow: first use an abstract proxy to craft the global trend, and then refine details with a more articulated model. We consider these feature suggestions highly inspiring and encouraging, and plan to explore them in future iterations.
\end{itemize}

\bibliographystyle{ACM-Reference-Format}

\begin{thebibliography}{62}
	
	
	\ifx \showCODEN    \undefined \def \showCODEN     #1{\unskip}     \fi
	\ifx \showISBNx    \undefined \def \showISBNx     #1{\unskip}     \fi
	\ifx \showISBNxiii \undefined \def \showISBNxiii  #1{\unskip}     \fi
	\ifx \showISSN     \undefined \def \showISSN      #1{\unskip}     \fi
	\ifx \showLCCN     \undefined \def \showLCCN      #1{\unskip}     \fi
	\ifx \shownote     \undefined \def \shownote      #1{#1}          \fi
	\ifx \showarticletitle \undefined \def \showarticletitle #1{#1}   \fi
	\ifx \showURL      \undefined \def \showURL       {\relax}        \fi
	\providecommand\bibfield[2]{#2}
	\providecommand\bibinfo[2]{#2}
	\providecommand\natexlab[1]{#1}
	\providecommand\showeprint[2][]{arXiv:#2}
	
	\bibitem[Ahuja et~al\mbox{.}(2021)]%
	{ahuja2021coolmoves}
	\bibfield{author}{\bibinfo{person}{Karan Ahuja}, \bibinfo{person}{Eyal Ofek}, \bibinfo{person}{Mar Gonzalez-Franco}, \bibinfo{person}{Christian Holz}, {and} \bibinfo{person}{Andrew~D Wilson}.} \bibinfo{year}{2021}\natexlab{}.
	\newblock \showarticletitle{Coolmoves: User motion accentuation in virtual reality}.
	\newblock \bibinfo{journal}{\emph{Proceedings of the ACM on Interactive, Mobile, Wearable and Ubiquitous Technologies}} \bibinfo{volume}{5}, \bibinfo{number}{2} (\bibinfo{year}{2021}), \bibinfo{pages}{1--23}.
	\newblock
	
	
	\bibitem[Awais et~al\mbox{.}(2025)]%
	{awais2025foundation}
	\bibfield{author}{\bibinfo{person}{Muhammad Awais}, \bibinfo{person}{Muzammal Naseer}, \bibinfo{person}{Salman Khan}, \bibinfo{person}{Rao~Muhammad Anwer}, \bibinfo{person}{Hisham Cholakkal}, \bibinfo{person}{Mubarak Shah}, \bibinfo{person}{Ming-Hsuan Yang}, {and} \bibinfo{person}{Fahad~Shahbaz Khan}.} \bibinfo{year}{2025}\natexlab{}.
	\newblock \showarticletitle{Foundation models defining a new era in vision: a survey and outlook}.
	\newblock \bibinfo{journal}{\emph{IEEE Transactions on Pattern Analysis and Machine Intelligence}} (\bibinfo{year}{2025}).
	\newblock
	
	
	\bibitem[Azadi et~al\mbox{.}(2023)]%
	{azadi2023make}
	\bibfield{author}{\bibinfo{person}{Samaneh Azadi}, \bibinfo{person}{Akbar Shah}, \bibinfo{person}{Thomas Hayes}, \bibinfo{person}{Devi Parikh}, {and} \bibinfo{person}{Sonal Gupta}.} \bibinfo{year}{2023}\natexlab{}.
	\newblock \showarticletitle{Make-an-animation: Large-scale text-conditional 3d human motion generation}. In \bibinfo{booktitle}{\emph{Proceedings of the IEEE/CVF International Conference on Computer Vision}}. \bibinfo{pages}{15039--15048}.
	\newblock
	
	
	\bibitem[Chen et~al\mbox{.}(2012)]%
	{chen2012kinetre}
	\bibfield{author}{\bibinfo{person}{Jiawen Chen}, \bibinfo{person}{Shahram Izadi}, {and} \bibinfo{person}{Andrew Fitzgibbon}.} \bibinfo{year}{2012}\natexlab{}.
	\newblock \showarticletitle{Kin{\^E}tre: animating the world with the human body}. In \bibinfo{booktitle}{\emph{Proceedings of the 25th annual ACM symposium on User interface software and technology}}. \bibinfo{pages}{435--444}.
	\newblock
	
	
	\bibitem[Croitoru et~al\mbox{.}(2023)]%
	{croitoru2023diffusion}
	\bibfield{author}{\bibinfo{person}{Florinel-Alin Croitoru}, \bibinfo{person}{Vlad Hondru}, \bibinfo{person}{Radu~Tudor Ionescu}, {and} \bibinfo{person}{Mubarak Shah}.} \bibinfo{year}{2023}\natexlab{}.
	\newblock \showarticletitle{Diffusion models in vision: A survey}.
	\newblock \bibinfo{journal}{\emph{IEEE transactions on pattern analysis and machine intelligence}} \bibinfo{volume}{45}, \bibinfo{number}{9} (\bibinfo{year}{2023}), \bibinfo{pages}{10850--10869}.
	\newblock
	
	
	\bibitem[Dai et~al\mbox{.}(2024)]%
	{dai2024motionlcm}
	\bibfield{author}{\bibinfo{person}{Wenxun Dai}, \bibinfo{person}{Ling-Hao Chen}, \bibinfo{person}{Jingbo Wang}, \bibinfo{person}{Jinpeng Liu}, \bibinfo{person}{Bo Dai}, {and} \bibinfo{person}{Yansong Tang}.} \bibinfo{year}{2024}\natexlab{}.
	\newblock \showarticletitle{Motionlcm: Real-time controllable motion generation via latent consistency model}. In \bibinfo{booktitle}{\emph{European Conference on Computer Vision}}. Springer, \bibinfo{pages}{390--408}.
	\newblock
	
	
	\bibitem[Desmarais et~al\mbox{.}(2021)]%
	{desmarais2021review}
	\bibfield{author}{\bibinfo{person}{Yann Desmarais}, \bibinfo{person}{Denis Mottet}, \bibinfo{person}{Pierre Slangen}, {and} \bibinfo{person}{Philippe Montesinos}.} \bibinfo{year}{2021}\natexlab{}.
	\newblock \showarticletitle{A review of 3D human pose estimation algorithms for markerless motion capture}.
	\newblock \bibinfo{journal}{\emph{Computer Vision and Image Understanding}}  \bibinfo{volume}{212} (\bibinfo{year}{2021}), \bibinfo{pages}{103275}.
	\newblock
	
	
	\bibitem[Dhariwal and Nichol(2021)]%
	{dhariwal2021diffusion}
	\bibfield{author}{\bibinfo{person}{Prafulla Dhariwal} {and} \bibinfo{person}{Alexander Nichol}.} \bibinfo{year}{2021}\natexlab{}.
	\newblock \showarticletitle{Diffusion models beat gans on image synthesis}.
	\newblock \bibinfo{journal}{\emph{Advances in neural information processing systems}}  \bibinfo{volume}{34} (\bibinfo{year}{2021}), \bibinfo{pages}{8780--8794}.
	\newblock
	
	
	\bibitem[Fang and Pollard(2003)]%
	{fang2003efficient}
	\bibfield{author}{\bibinfo{person}{Anthony~C Fang} {and} \bibinfo{person}{Nancy~S Pollard}.} \bibinfo{year}{2003}\natexlab{}.
	\newblock \showarticletitle{Efficient synthesis of physically valid human motion}.
	\newblock \bibinfo{journal}{\emph{ACM Transactions on Graphics (TOG)}} \bibinfo{volume}{22}, \bibinfo{number}{3} (\bibinfo{year}{2003}), \bibinfo{pages}{417--426}.
	\newblock
	
	
	\bibitem[Gat et~al\mbox{.}(2025)]%
	{gat2025anytop}
	\bibfield{author}{\bibinfo{person}{Inbar Gat}, \bibinfo{person}{Sigal Raab}, \bibinfo{person}{Guy Tevet}, \bibinfo{person}{Yuval Reshef}, \bibinfo{person}{Amit~Haim Bermano}, {and} \bibinfo{person}{Daniel Cohen-Or}.} \bibinfo{year}{2025}\natexlab{}.
	\newblock \showarticletitle{Anytop: Character animation diffusion with any topology}. In \bibinfo{booktitle}{\emph{Proceedings of the Special Interest Group on Computer Graphics and Interactive Techniques Conference Conference Papers}}. \bibinfo{pages}{1--10}.
	\newblock
	
	
	\bibitem[Glauser et~al\mbox{.}(2016)]%
	{glauser2016rig}
	\bibfield{author}{\bibinfo{person}{Oliver Glauser}, \bibinfo{person}{Wan-Chun Ma}, \bibinfo{person}{Daniele Panozzo}, \bibinfo{person}{Alec Jacobson}, \bibinfo{person}{Otmar Hilliges}, {and} \bibinfo{person}{Olga Sorkine-Hornung}.} \bibinfo{year}{2016}\natexlab{}.
	\newblock \showarticletitle{Rig animation with a tangible and modular input device}.
	\newblock \bibinfo{journal}{\emph{ACM Transactions on Graphics (TOG)}} \bibinfo{volume}{35}, \bibinfo{number}{4} (\bibinfo{year}{2016}), \bibinfo{pages}{1--11}.
	\newblock
	
	
	\bibitem[Goodfellow et~al\mbox{.}(2014)]%
	{goodfellow2014generative}
	\bibfield{author}{\bibinfo{person}{Ian~J Goodfellow}, \bibinfo{person}{Jean Pouget-Abadie}, \bibinfo{person}{Mehdi Mirza}, \bibinfo{person}{Bing Xu}, \bibinfo{person}{David Warde-Farley}, \bibinfo{person}{Sherjil Ozair}, \bibinfo{person}{Aaron Courville}, {and} \bibinfo{person}{Yoshua Bengio}.} \bibinfo{year}{2014}\natexlab{}.
	\newblock \showarticletitle{Generative adversarial nets}.
	\newblock \bibinfo{journal}{\emph{Advances in neural information processing systems}}  \bibinfo{volume}{27} (\bibinfo{year}{2014}).
	\newblock
	
	
	\bibitem[Guo et~al\mbox{.}(2024)]%
	{guo2024momask}
	\bibfield{author}{\bibinfo{person}{Chuan Guo}, \bibinfo{person}{Yuxuan Mu}, \bibinfo{person}{Muhammad~Gohar Javed}, \bibinfo{person}{Sen Wang}, {and} \bibinfo{person}{Li Cheng}.} \bibinfo{year}{2024}\natexlab{}.
	\newblock \showarticletitle{Momask: Generative masked modeling of 3d human motions}. In \bibinfo{booktitle}{\emph{Proceedings of the IEEE/CVF Conference on Computer Vision and Pattern Recognition}}. \bibinfo{pages}{1900--1910}.
	\newblock
	
	
	\bibitem[Guo et~al\mbox{.}(2022)]%
	{Guo_2022_CVPR}
	\bibfield{author}{\bibinfo{person}{Chuan Guo}, \bibinfo{person}{Shihao Zou}, \bibinfo{person}{Xinxin Zuo}, \bibinfo{person}{Sen Wang}, \bibinfo{person}{Wei Ji}, \bibinfo{person}{Xingyu Li}, {and} \bibinfo{person}{Li Cheng}.} \bibinfo{year}{2022}\natexlab{}.
	\newblock \showarticletitle{Generating Diverse and Natural 3D Human Motions From Text}. In \bibinfo{booktitle}{\emph{Proceedings of the IEEE/CVF Conference on Computer Vision and Pattern Recognition (CVPR)}}. \bibinfo{pages}{5152--5161}.
	\newblock
	
	
	\bibitem[Gupta et~al\mbox{.}(2014)]%
	{gupta2014motionmontage}
	\bibfield{author}{\bibinfo{person}{Ankit Gupta}, \bibinfo{person}{Maneesh Agrawala}, \bibinfo{person}{Brian Curless}, {and} \bibinfo{person}{Michael Cohen}.} \bibinfo{year}{2014}\natexlab{}.
	\newblock \showarticletitle{Motionmontage: A system to annotate and combine motion takes for 3d animations}. In \bibinfo{booktitle}{\emph{Proceedings of the SIGCHI Conference on Human Factors in Computing Systems}}. \bibinfo{pages}{2017--2026}.
	\newblock
	
	
	\bibitem[Held et~al\mbox{.}(2012)]%
	{held20123dkinect}
	\bibfield{author}{\bibinfo{person}{Robert Held}, \bibinfo{person}{Ankit Gupta}, \bibinfo{person}{Brian Curless}, {and} \bibinfo{person}{Maneesh Agrawala}.} \bibinfo{year}{2012}\natexlab{}.
	\newblock \showarticletitle{3D puppetry: a kinect-based interface for 3D animation.}. In \bibinfo{booktitle}{\emph{UIST}}, Vol.~\bibinfo{volume}{12}. \bibinfo{pages}{423--434}.
	\newblock
	
	
	\bibitem[Ho and Salimans(2022)]%
	{ho2022classifier}
	\bibfield{author}{\bibinfo{person}{Jonathan Ho} {and} \bibinfo{person}{Tim Salimans}.} \bibinfo{year}{2022}\natexlab{}.
	\newblock \showarticletitle{Classifier-free diffusion guidance}.
	\newblock \bibinfo{journal}{\emph{arXiv preprint arXiv:2207.12598}} (\bibinfo{year}{2022}).
	\newblock
	
	
	\bibitem[Hung et~al\mbox{.}(2022)]%
	{hung2022puppeteer}
	\bibfield{author}{\bibinfo{person}{Ching-Wen Hung}, \bibinfo{person}{Ruei-Che Chang}, \bibinfo{person}{Hong-Sheng Chen}, \bibinfo{person}{Chung~Han Liang}, \bibinfo{person}{Liwei Chan}, {and} \bibinfo{person}{Bing-Yu Chen}.} \bibinfo{year}{2022}\natexlab{}.
	\newblock \showarticletitle{Puppeteer: Exploring intuitive hand gestures and upper-body postures for manipulating human avatar actions}. In \bibinfo{booktitle}{\emph{proceedings of the 28th ACM symposium on virtual reality software and technology}}. \bibinfo{pages}{1--11}.
	\newblock
	
	
	\bibitem[Hung et~al\mbox{.}(2024)]%
	{hung2024fingerpuppet}
	\bibfield{author}{\bibinfo{person}{Ching-Wen Hung}, \bibinfo{person}{Chung-Han Liang}, {and} \bibinfo{person}{Bing-Yu Chen}.} \bibinfo{year}{2024}\natexlab{}.
	\newblock \showarticletitle{Fingerpuppet: finger-walking performance-based puppetry for human avatar}. In \bibinfo{booktitle}{\emph{Extended Abstracts of the CHI Conference on Human Factors in Computing Systems}}. \bibinfo{pages}{1--6}.
	\newblock
	
	
	\bibitem[Jacobson et~al\mbox{.}(2014)]%
	{jacobson2014tangible}
	\bibfield{author}{\bibinfo{person}{Alec Jacobson}, \bibinfo{person}{Daniele Panozzo}, \bibinfo{person}{Oliver Glauser}, \bibinfo{person}{C{\'e}dric Pradalier}, \bibinfo{person}{Otmar Hilliges}, {and} \bibinfo{person}{Olga Sorkine-Hornung}.} \bibinfo{year}{2014}\natexlab{}.
	\newblock \showarticletitle{Tangible and modular input device for character articulation}.
	\newblock \bibinfo{journal}{\emph{ACM Transactions on Graphics (TOG)}} \bibinfo{volume}{33}, \bibinfo{number}{4} (\bibinfo{year}{2014}), \bibinfo{pages}{1--12}.
	\newblock
	
	
	\bibitem[Jiang et~al\mbox{.}(2023)]%
	{jiang2023handavatar}
	\bibfield{author}{\bibinfo{person}{Yu Jiang}, \bibinfo{person}{Zhipeng Li}, \bibinfo{person}{Mufei He}, \bibinfo{person}{David Lindlbauer}, {and} \bibinfo{person}{Yukang Yan}.} \bibinfo{year}{2023}\natexlab{}.
	\newblock \showarticletitle{Handavatar: Embodying non-humanoid virtual avatars through hands}. In \bibinfo{booktitle}{\emph{Proceedings of the 2023 CHI conference on human factors in computing systems}}. \bibinfo{pages}{1--17}.
	\newblock
	
	
	\bibitem[Karaev et~al\mbox{.}(2024)]%
	{karaev2024cotracker3}
	\bibfield{author}{\bibinfo{person}{Nikita Karaev}, \bibinfo{person}{Iurii Makarov}, \bibinfo{person}{Jianyuan Wang}, \bibinfo{person}{Natalia Neverova}, \bibinfo{person}{Andrea Vedaldi}, {and} \bibinfo{person}{Christian Rupprecht}.} \bibinfo{year}{2024}\natexlab{}.
	\newblock \showarticletitle{Cotracker3: Simpler and better point tracking by pseudo-labelling real videos}.
	\newblock \bibinfo{journal}{\emph{arXiv preprint arXiv:2410.11831}} (\bibinfo{year}{2024}).
	\newblock
	
	
	\bibitem[Kingma and Welling(2013)]%
	{kingma2013auto}
	\bibfield{author}{\bibinfo{person}{Diederik~P Kingma} {and} \bibinfo{person}{Max Welling}.} \bibinfo{year}{2013}\natexlab{}.
	\newblock \showarticletitle{Auto-encoding variational bayes}.
	\newblock \bibinfo{journal}{\emph{arXiv preprint arXiv:1312.6114}} (\bibinfo{year}{2013}).
	\newblock
	
	
	\bibitem[Kirillov et~al\mbox{.}(2023)]%
	{kirillov2023segment}
	\bibfield{author}{\bibinfo{person}{Alexander Kirillov}, \bibinfo{person}{Eric Mintun}, \bibinfo{person}{Nikhila Ravi}, \bibinfo{person}{Hanzi Mao}, \bibinfo{person}{Chloe Rolland}, \bibinfo{person}{Laura Gustafson}, \bibinfo{person}{Tete Xiao}, \bibinfo{person}{Spencer Whitehead}, \bibinfo{person}{Alexander~C Berg}, \bibinfo{person}{Wan-Yen Lo}, {et~al\mbox{.}}} \bibinfo{year}{2023}\natexlab{}.
	\newblock \showarticletitle{Segment anything}. In \bibinfo{booktitle}{\emph{Proceedings of the IEEE/CVF international conference on computer vision}}. \bibinfo{pages}{4015--4026}.
	\newblock
	
	
	\bibitem[Knep et~al\mbox{.}(1995)]%
	{Dinosaur1995}
	\bibfield{author}{\bibinfo{person}{Brian Knep}, \bibinfo{person}{Craig Hayes}, \bibinfo{person}{Rick Sayre}, {and} \bibinfo{person}{Tom Williams}.} \bibinfo{year}{1995}\natexlab{}.
	\newblock \showarticletitle{Dinosaur input device}. In \bibinfo{booktitle}{\emph{ACM Conference on Human Factors in Computing Systems (CHI)}}.
	\newblock
	
	
	\bibitem[Lamberti et~al\mbox{.}(2017)]%
	{lamberti2017virtual}
	\bibfield{author}{\bibinfo{person}{Fabrizio Lamberti}, \bibinfo{person}{Gianluca Paravati}, \bibinfo{person}{Valentina Gatteschi}, \bibinfo{person}{Alberto Cannavo}, {and} \bibinfo{person}{Paolo Montuschi}.} \bibinfo{year}{2017}\natexlab{}.
	\newblock \showarticletitle{Virtual character animation based on affordable motion capture and reconfigurable tangible interfaces}.
	\newblock \bibinfo{journal}{\emph{IEEE transactions on visualization and computer graphics}} \bibinfo{volume}{24}, \bibinfo{number}{5} (\bibinfo{year}{2017}), \bibinfo{pages}{1742--1755}.
	\newblock
	
	
	\bibitem[Li et~al\mbox{.}(2024)]%
	{li2024anicraft}
	\bibfield{author}{\bibinfo{person}{Boyu Li}, \bibinfo{person}{Linping Yuan}, \bibinfo{person}{Zhe Yan}, \bibinfo{person}{Qianxi Liu}, \bibinfo{person}{Yulin Shen}, {and} \bibinfo{person}{Zeyu Wang}.} \bibinfo{year}{2024}\natexlab{}.
	\newblock \showarticletitle{Anicraft: Crafting everyday objects as physical proxies for prototyping 3d character animation in mixed reality}. In \bibinfo{booktitle}{\emph{Proceedings of the 37th Annual ACM Symposium on User Interface Software and Technology}}. \bibinfo{pages}{1--14}.
	\newblock
	
	
	\bibitem[Li et~al\mbox{.}(2025)]%
	{li2025genmo}
	\bibfield{author}{\bibinfo{person}{Jiefeng Li}, \bibinfo{person}{Jinkun Cao}, \bibinfo{person}{Haotian Zhang}, \bibinfo{person}{Davis Rempe}, \bibinfo{person}{Jan Kautz}, \bibinfo{person}{Umar Iqbal}, {and} \bibinfo{person}{Ye Yuan}.} \bibinfo{year}{2025}\natexlab{}.
	\newblock \showarticletitle{GENMO: A GENeralist Model for Human MOtion}.
	\newblock \bibinfo{journal}{\emph{arXiv preprint arXiv:2505.01425}} (\bibinfo{year}{2025}).
	\newblock
	
	
	\bibitem[Lipman et~al\mbox{.}(2022)]%
	{lipman2022flow}
	\bibfield{author}{\bibinfo{person}{Yaron Lipman}, \bibinfo{person}{Ricky~TQ Chen}, \bibinfo{person}{Heli Ben-Hamu}, \bibinfo{person}{Maximilian Nickel}, {and} \bibinfo{person}{Matt Le}.} \bibinfo{year}{2022}\natexlab{}.
	\newblock \showarticletitle{Flow matching for generative modeling}.
	\newblock \bibinfo{journal}{\emph{arXiv preprint arXiv:2210.02747}} (\bibinfo{year}{2022}).
	\newblock
	
	
	\bibitem[Liu et~al\mbox{.}(2024)]%
	{SketchDream}
	\bibfield{author}{\bibinfo{person}{Feng-Lin Liu}, \bibinfo{person}{Hongbo Fu}, \bibinfo{person}{Yu-Kun Lai}, {and} \bibinfo{person}{Lin Gao}.} \bibinfo{year}{2024}\natexlab{}.
	\newblock \showarticletitle{SketchDream: Sketch-based Text-To-3D Generation and Editing}.
	\newblock \bibinfo{journal}{\emph{ACM Transactions on Graphics (Proc. SIGGRAPH)}} \bibinfo{volume}{43}, \bibinfo{number}{4} (\bibinfo{year}{2024}).
	\newblock
	\showISSN{0730-0301}
	\href{https://doi.org/10.1145/3658120}{doi:\nolinkurl{10.1145/3658120}}
	
	
	\bibitem[Lockwood and Singh(2012)]%
	{lockwood2012fingerwalking}
	\bibfield{author}{\bibinfo{person}{Noah Lockwood} {and} \bibinfo{person}{Karan Singh}.} \bibinfo{year}{2012}\natexlab{}.
	\newblock \showarticletitle{Fingerwalking: motion editing with contact-based hand performance}. In \bibinfo{booktitle}{\emph{Proceedings of the 11th ACM SIGGRAPH/Eurographics conference on Computer Animation}}. \bibinfo{pages}{43--52}.
	\newblock
	
	
	\bibitem[Mao et~al\mbox{.}(2023)]%
	{mao20233d}
	\bibfield{author}{\bibinfo{person}{Jiageng Mao}, \bibinfo{person}{Shaoshuai Shi}, \bibinfo{person}{Xiaogang Wang}, {and} \bibinfo{person}{Hongsheng Li}.} \bibinfo{year}{2023}\natexlab{}.
	\newblock \showarticletitle{3D object detection for autonomous driving: A comprehensive survey}.
	\newblock \bibinfo{journal}{\emph{International Journal of Computer Vision}} \bibinfo{volume}{131}, \bibinfo{number}{8} (\bibinfo{year}{2023}), \bibinfo{pages}{1909--1963}.
	\newblock
	
	
	\bibitem[Moeslund et~al\mbox{.}(2006)]%
	{moeslund2006survey}
	\bibfield{author}{\bibinfo{person}{Thomas~B Moeslund}, \bibinfo{person}{Adrian Hilton}, {and} \bibinfo{person}{Volker Kr{\"u}ger}.} \bibinfo{year}{2006}\natexlab{}.
	\newblock \showarticletitle{A survey of advances in vision-based human motion capture and analysis}.
	\newblock \bibinfo{journal}{\emph{Computer vision and image understanding}} \bibinfo{volume}{104}, \bibinfo{number}{2-3} (\bibinfo{year}{2006}), \bibinfo{pages}{90--126}.
	\newblock
	
	
	\bibitem[Numaguchi et~al\mbox{.}(2011)]%
	{Numaguchi2011}
	\bibfield{author}{\bibinfo{person}{Naoki Numaguchi}, \bibinfo{person}{Atsushi Nakazawa}, \bibinfo{person}{Takaaki Shiratori}, {and} \bibinfo{person}{Jessica~K. Hodgins}.} \bibinfo{year}{2011}\natexlab{}.
	\newblock \showarticletitle{A puppet interface for retrieval of motion capture data}. In \bibinfo{booktitle}{\emph{ACM SIGGRAPH/Eurographics Symposium on Computer Animation}}.
	\newblock
	
	
	\bibitem[Park and Ravani(1997)]%
	{park1997smooth}
	\bibfield{author}{\bibinfo{person}{Frank~C Park} {and} \bibinfo{person}{Bahram Ravani}.} \bibinfo{year}{1997}\natexlab{}.
	\newblock \showarticletitle{Smooth invariant interpolation of rotations}.
	\newblock \bibinfo{journal}{\emph{ACM Transactions on Graphics (TOG)}} \bibinfo{volume}{16}, \bibinfo{number}{3} (\bibinfo{year}{1997}), \bibinfo{pages}{277--295}.
	\newblock
	
	
	\bibitem[Peng et~al\mbox{.}(2023)]%
	{peng2023dualmotion}
	\bibfield{author}{\bibinfo{person}{Yichen Peng}, \bibinfo{person}{Chunqi Zhao}, \bibinfo{person}{Haoran Xie}, \bibinfo{person}{Tsukasa Fukusato}, \bibinfo{person}{Kazunori Miyata}, {and} \bibinfo{person}{Takeo Igarashi}.} \bibinfo{year}{2023}\natexlab{}.
	\newblock \showarticletitle{Dualmotion: Global-to-local casual motion design for character animations}.
	\newblock \bibinfo{journal}{\emph{IEICE TRANSACTIONS on Information and Systems}} \bibinfo{volume}{106}, \bibinfo{number}{4} (\bibinfo{year}{2023}), \bibinfo{pages}{459--468}.
	\newblock
	
	
	\bibitem[Plappert et~al\mbox{.}(2016)]%
	{Plappert2016}
	\bibfield{author}{\bibinfo{person}{Matthias Plappert}, \bibinfo{person}{Christian Mandery}, {and} \bibinfo{person}{Tamim Asfour}.} \bibinfo{year}{2016}\natexlab{}.
	\newblock \showarticletitle{The {KIT} Motion-Language Dataset}.
	\newblock \bibinfo{journal}{\emph{Big Data}} \bibinfo{volume}{4}, \bibinfo{number}{4} (\bibinfo{date}{dec} \bibinfo{year}{2016}), \bibinfo{pages}{236--252}.
	\newblock
	\href{https://doi.org/10.1089/big.2016.0028}{doi:\nolinkurl{10.1089/big.2016.0028}}
	
	
	\bibitem[Qi et~al\mbox{.}(2017)]%
	{qi2017pointnet}
	\bibfield{author}{\bibinfo{person}{Charles~R Qi}, \bibinfo{person}{Hao Su}, \bibinfo{person}{Kaichun Mo}, {and} \bibinfo{person}{Leonidas~J Guibas}.} \bibinfo{year}{2017}\natexlab{}.
	\newblock \showarticletitle{Pointnet: Deep learning on point sets for 3d classification and segmentation}. In \bibinfo{booktitle}{\emph{Proceedings of the IEEE conference on computer vision and pattern recognition}}. \bibinfo{pages}{652--660}.
	\newblock
	
	
	\bibitem[Ravi et~al\mbox{.}(2024)]%
	{ravi2024sam}
	\bibfield{author}{\bibinfo{person}{Nikhila Ravi}, \bibinfo{person}{Valentin Gabeur}, \bibinfo{person}{Yuan-Ting Hu}, \bibinfo{person}{Ronghang Hu}, \bibinfo{person}{Chaitanya Ryali}, \bibinfo{person}{Tengyu Ma}, \bibinfo{person}{Haitham Khedr}, \bibinfo{person}{Roman R{\"a}dle}, \bibinfo{person}{Chloe Rolland}, \bibinfo{person}{Laura Gustafson}, {et~al\mbox{.}}} \bibinfo{year}{2024}\natexlab{}.
	\newblock \showarticletitle{Sam 2: Segment anything in images and videos}.
	\newblock \bibinfo{journal}{\emph{arXiv preprint arXiv:2408.00714}} (\bibinfo{year}{2024}).
	\newblock
	
	
	\bibitem[Rhodin et~al\mbox{.}(2015)]%
	{rhodin2015generalizing}
	\bibfield{author}{\bibinfo{person}{Helge Rhodin}, \bibinfo{person}{James Tompkin}, \bibinfo{person}{Kwang~In Kim}, \bibinfo{person}{Edilson De~Aguiar}, \bibinfo{person}{Hanspeter Pfister}, \bibinfo{person}{Hans-Peter Seidel}, {and} \bibinfo{person}{Christian Theobalt}.} \bibinfo{year}{2015}\natexlab{}.
	\newblock \showarticletitle{Generalizing wave gestures from sparse examples for real-time character control}.
	\newblock \bibinfo{journal}{\emph{ACM Transactions on Graphics (TOG)}} \bibinfo{volume}{34}, \bibinfo{number}{6} (\bibinfo{year}{2015}), \bibinfo{pages}{1--12}.
	\newblock
	
	
	\bibitem[Saint-Aubert et~al\mbox{.}(2023)]%
	{saint2023tangible}
	\bibfield{author}{\bibinfo{person}{Justine Saint-Aubert}, \bibinfo{person}{Ferran Argelaguet}, {and} \bibinfo{person}{Anatole L{\'e}cuyer}.} \bibinfo{year}{2023}\natexlab{}.
	\newblock \showarticletitle{Tangible Avatar: Enhancing Presence and Embodiment During Seated Virtual Experiences with a Prop-Based Controller}. In \bibinfo{booktitle}{\emph{2023 IEEE International Symposium on Mixed and Augmented Reality Adjunct (ISMAR-Adjunct)}}. IEEE, \bibinfo{pages}{572--577}.
	\newblock
	
	
	\bibitem[Shimada et~al\mbox{.}(2020)]%
	{shimada2020physcap}
	\bibfield{author}{\bibinfo{person}{Soshi Shimada}, \bibinfo{person}{Vladislav Golyanik}, \bibinfo{person}{Weipeng Xu}, {and} \bibinfo{person}{Christian Theobalt}.} \bibinfo{year}{2020}\natexlab{}.
	\newblock \showarticletitle{Physcap: Physically plausible monocular 3d motion capture in real time}.
	\newblock \bibinfo{journal}{\emph{ACM Transactions on Graphics (ToG)}} \bibinfo{volume}{39}, \bibinfo{number}{6} (\bibinfo{year}{2020}), \bibinfo{pages}{1--16}.
	\newblock
	
	
	\bibitem[Sin et~al\mbox{.}(2022)]%
	{sin2022tracking}
	\bibfield{author}{\bibinfo{person}{Zackary~PT Sin}, \bibinfo{person}{Peter~Q Chen}, \bibinfo{person}{Peter~HF Ng}, {and} \bibinfo{person}{Hong~Va Leong}.} \bibinfo{year}{2022}\natexlab{}.
	\newblock \showarticletitle{Tracking stuffed toy for naturally mapped interactive play via a soft-pose estimator}.
	\newblock \bibinfo{journal}{\emph{Proceedings of the ACM on Human-Computer Interaction}} \bibinfo{volume}{6}, \bibinfo{number}{CHI PLAY} (\bibinfo{year}{2022}), \bibinfo{pages}{1--25}.
	\newblock
	
	
	\bibitem[Song et~al\mbox{.}(2025)]%
	{song2025puppeteer}
	\bibfield{author}{\bibinfo{person}{Chaoyue Song}, \bibinfo{person}{Xiu Li}, \bibinfo{person}{Fan Yang}, \bibinfo{person}{Zhongcong Xu}, \bibinfo{person}{Jiacheng Wei}, \bibinfo{person}{Fayao Liu}, \bibinfo{person}{Jiashi Feng}, \bibinfo{person}{Guosheng Lin}, {and} \bibinfo{person}{Jianfeng Zhang}.} \bibinfo{year}{2025}\natexlab{}.
	\newblock \showarticletitle{Puppeteer: Rig and Animate Your 3D Models}.
	\newblock \bibinfo{journal}{\emph{arXiv preprint arXiv:2508.10898}} (\bibinfo{year}{2025}).
	\newblock
	
	
	\bibitem[Song et~al\mbox{.}(2023)]%
	{song2023loss}
	\bibfield{author}{\bibinfo{person}{Jiaming Song}, \bibinfo{person}{Qinsheng Zhang}, \bibinfo{person}{Hongxu Yin}, \bibinfo{person}{Morteza Mardani}, \bibinfo{person}{Ming-Yu Liu}, \bibinfo{person}{Jan Kautz}, \bibinfo{person}{Yongxin Chen}, {and} \bibinfo{person}{Arash Vahdat}.} \bibinfo{year}{2023}\natexlab{}.
	\newblock \showarticletitle{Loss-guided diffusion models for plug-and-play controllable generation}. In \bibinfo{booktitle}{\emph{International Conference on Machine Learning}}. PMLR, \bibinfo{pages}{32483--32498}.
	\newblock
	
	
	\bibitem[Tevet et~al\mbox{.}(2022)]%
	{tevet2022human}
	\bibfield{author}{\bibinfo{person}{Guy Tevet}, \bibinfo{person}{Sigal Raab}, \bibinfo{person}{Brian Gordon}, \bibinfo{person}{Yonatan Shafir}, \bibinfo{person}{Daniel Cohen-Or}, {and} \bibinfo{person}{Amit~H Bermano}.} \bibinfo{year}{2022}\natexlab{}.
	\newblock \showarticletitle{Human motion diffusion model}.
	\newblock \bibinfo{journal}{\emph{arXiv preprint arXiv:2209.14916}} (\bibinfo{year}{2022}).
	\newblock
	
	
	\bibitem[Umeyama(2002)]%
	{umeyama2002least}
	\bibfield{author}{\bibinfo{person}{Shinji Umeyama}.} \bibinfo{year}{2002}\natexlab{}.
	\newblock \showarticletitle{Least-squares estimation of transformation parameters between two point patterns}.
	\newblock \bibinfo{journal}{\emph{IEEE Transactions on pattern analysis and machine intelligence}} \bibinfo{volume}{13}, \bibinfo{number}{4} (\bibinfo{year}{2002}), \bibinfo{pages}{376--380}.
	\newblock
	
	
	\bibitem[Wang et~al\mbox{.}(2024)]%
	{wang2024egocentric}
	\bibfield{author}{\bibinfo{person}{Jian Wang}, \bibinfo{person}{Zhe Cao}, \bibinfo{person}{Diogo Luvizon}, \bibinfo{person}{Lingjie Liu}, \bibinfo{person}{Kripasindhu Sarkar}, \bibinfo{person}{Danhang Tang}, \bibinfo{person}{Thabo Beeler}, {and} \bibinfo{person}{Christian Theobalt}.} \bibinfo{year}{2024}\natexlab{}.
	\newblock \showarticletitle{Egocentric whole-body motion capture with fisheyevit and diffusion-based motion refinement}. In \bibinfo{booktitle}{\emph{Proceedings of the IEEE/CVF Conference on Computer Vision and Pattern Recognition}}. \bibinfo{pages}{777--787}.
	\newblock
	
	
	\bibitem[Wang et~al\mbox{.}(2025a)]%
	{wang2025vggt}
	\bibfield{author}{\bibinfo{person}{Jianyuan Wang}, \bibinfo{person}{Minghao Chen}, \bibinfo{person}{Nikita Karaev}, \bibinfo{person}{Andrea Vedaldi}, \bibinfo{person}{Christian Rupprecht}, {and} \bibinfo{person}{David Novotny}.} \bibinfo{year}{2025}\natexlab{a}.
	\newblock \showarticletitle{Vggt: Visual geometry grounded transformer}. In \bibinfo{booktitle}{\emph{Proceedings of the Computer Vision and Pattern Recognition Conference}}. \bibinfo{pages}{5294--5306}.
	\newblock
	
	
	\bibitem[Wang et~al\mbox{.}(2025b)]%
	{wang2025x-mogen}
	\bibfield{author}{\bibinfo{person}{Xuan Wang}, \bibinfo{person}{Kai Ruan}, \bibinfo{person}{Liyang Qian}, \bibinfo{person}{Zhizhi Guo}, \bibinfo{person}{Chang Su}, {and} \bibinfo{person}{Gaoang Wang}.} \bibinfo{year}{2025}\natexlab{b}.
	\newblock \showarticletitle{X-MoGen: Unified Motion Generation across Humans and Animals}.
	\newblock \bibinfo{journal}{\emph{arXiv preprint arXiv:2508.05162}} (\bibinfo{year}{2025}).
	\newblock
	
	
	\bibitem[Wang et~al\mbox{.}(2025c)]%
	{wang2025pi}
	\bibfield{author}{\bibinfo{person}{Yifan Wang}, \bibinfo{person}{Jianjun Zhou}, \bibinfo{person}{Haoyi Zhu}, \bibinfo{person}{Wenzheng Chang}, \bibinfo{person}{Yang Zhou}, \bibinfo{person}{Zizun Li}, \bibinfo{person}{Junyi Chen}, \bibinfo{person}{Jiangmiao Pang}, \bibinfo{person}{Chunhua Shen}, {and} \bibinfo{person}{Tong He}.} \bibinfo{year}{2025}\natexlab{c}.
	\newblock \showarticletitle{\(\pi^3\): Scalable Permutation-Equivariant Visual Geometry Learning}.
	\newblock \bibinfo{journal}{\emph{arXiv preprint arXiv:2507.13347}} (\bibinfo{year}{2025}).
	\newblock
	
	
	\bibitem[Wang et~al\mbox{.}(2023)]%
	{wang2023intercontrol}
	\bibfield{author}{\bibinfo{person}{Zhenzhi Wang}, \bibinfo{person}{Jingbo Wang}, \bibinfo{person}{Dahua Lin}, {and} \bibinfo{person}{Bo Dai}.} \bibinfo{year}{2023}\natexlab{}.
	\newblock \showarticletitle{Intercontrol: Generate human motion interactions by controlling every joint}.
	\newblock \bibinfo{journal}{\emph{arXiv preprint arXiv:2311.15864}}  \bibinfo{volume}{3} (\bibinfo{year}{2023}).
	\newblock
	
	
	\bibitem[Xie et~al\mbox{.}(2023)]%
	{xie2023omnicontrol}
	\bibfield{author}{\bibinfo{person}{Yiming Xie}, \bibinfo{person}{Varun Jampani}, \bibinfo{person}{Lei Zhong}, \bibinfo{person}{Deqing Sun}, {and} \bibinfo{person}{Huaizu Jiang}.} \bibinfo{year}{2023}\natexlab{}.
	\newblock \showarticletitle{Omnicontrol: Control any joint at any time for human motion generation}.
	\newblock \bibinfo{journal}{\emph{arXiv preprint arXiv:2310.08580}} (\bibinfo{year}{2023}).
	\newblock
	
	
	\bibitem[Xu et~al\mbox{.}(2025a)]%
	{xu2025mospa}
	\bibfield{author}{\bibinfo{person}{Shuyang Xu}, \bibinfo{person}{Zhiyang Dou}, \bibinfo{person}{Mingyi Shi}, \bibinfo{person}{Liang Pan}, \bibinfo{person}{Leo Ho}, \bibinfo{person}{Jingbo Wang}, \bibinfo{person}{Yuan Liu}, \bibinfo{person}{Cheng Lin}, \bibinfo{person}{Yuexin Ma}, \bibinfo{person}{Wenping Wang}, {et~al\mbox{.}}} \bibinfo{year}{2025}\natexlab{a}.
	\newblock \showarticletitle{MOSPA: Human Motion Generation Driven by Spatial Audio}.
	\newblock \bibinfo{journal}{\emph{arXiv preprint arXiv:2507.11949}} (\bibinfo{year}{2025}).
	\newblock
	
	
	\bibitem[Xu et~al\mbox{.}(2025b)]%
	{MultiPersonGen25}
	\bibfield{author}{\bibinfo{person}{Wenning Xu}, \bibinfo{person}{Shiyu Fan}, \bibinfo{person}{Paul Henderson}, {and} \bibinfo{person}{Edmond S.~L. Ho}.} \bibinfo{year}{2025}\natexlab{b}.
	\newblock \showarticletitle{Multi-Person Interaction Generation from Two-Person Motion Priors}. In \bibinfo{booktitle}{\emph{Proceedings of the Special Interest Group on Computer Graphics and Interactive Techniques Conference Conference Papers}} \emph{(\bibinfo{series}{SIGGRAPH Conference Papers '25})}. \bibinfo{publisher}{Association for Computing Machinery}, \bibinfo{address}{New York, NY, USA}, Article \bibinfo{articleno}{113}, \bibinfo{numpages}{11}~pages.
	\newblock
	\showISBNx{9798400715402}
	
	
	\bibitem[Yang et~al\mbox{.}(2024)]%
	{yang2024omnimotiongpt}
	\bibfield{author}{\bibinfo{person}{Zhangsihao Yang}, \bibinfo{person}{Mingyuan Zhou}, \bibinfo{person}{Mengyi Shan}, \bibinfo{person}{Bingbing Wen}, \bibinfo{person}{Ziwei Xuan}, \bibinfo{person}{Mitch Hill}, \bibinfo{person}{Junjie Bai}, \bibinfo{person}{Guo-Jun Qi}, {and} \bibinfo{person}{Yalin Wang}.} \bibinfo{year}{2024}\natexlab{}.
	\newblock \showarticletitle{Omnimotiongpt: Animal motion generation with limited data}. In \bibinfo{booktitle}{\emph{Proceedings of the IEEE/CVF Conference on Computer Vision and Pattern Recognition}}. \bibinfo{pages}{1249--1259}.
	\newblock
	
	
	\bibitem[Ye et~al\mbox{.}(2020)]%
	{ye2020aranimator}
	\bibfield{author}{\bibinfo{person}{Hui Ye}, \bibinfo{person}{Kin~Chung Kwan}, \bibinfo{person}{Wanchao Su}, {and} \bibinfo{person}{Hongbo Fu}.} \bibinfo{year}{2020}\natexlab{}.
	\newblock \showarticletitle{ARAnimator: In-situ character animation in mobile AR with user-defined motion gestures}.
	\newblock \bibinfo{journal}{\emph{ACM Transactions on Graphics (TOG)}} \bibinfo{volume}{39}, \bibinfo{number}{4} (\bibinfo{year}{2020}), \bibinfo{pages}{83--1}.
	\newblock
	
	
	\bibitem[Yoshizaki et~al\mbox{.}(2011)]%
	{yoshizaki2011actuated}
	\bibfield{author}{\bibinfo{person}{Wataru Yoshizaki}, \bibinfo{person}{Yuta Sugiura}, \bibinfo{person}{Albert~C Chiou}, \bibinfo{person}{Sunao Hashimoto}, \bibinfo{person}{Masahiko Inami}, \bibinfo{person}{Takeo Igarashi}, \bibinfo{person}{Yoshiaki Akazawa}, \bibinfo{person}{Katsuaki Kawachi}, \bibinfo{person}{Satoshi Kagami}, {and} \bibinfo{person}{Masaaki Mochimaru}.} \bibinfo{year}{2011}\natexlab{}.
	\newblock \showarticletitle{An actuated physical puppet as an input device for controlling a digital manikin}. In \bibinfo{booktitle}{\emph{Proceedings of the SIGCHI Conference on Human Factors in Computing Systems}}. \bibinfo{pages}{637--646}.
	\newblock
	
	
	\bibitem[Zhang et~al\mbox{.}(2023)]%
	{zhang2023adding}
	\bibfield{author}{\bibinfo{person}{Lvmin Zhang}, \bibinfo{person}{Anyi Rao}, {and} \bibinfo{person}{Maneesh Agrawala}.} \bibinfo{year}{2023}\natexlab{}.
	\newblock \showarticletitle{Adding conditional control to text-to-image diffusion models}. In \bibinfo{booktitle}{\emph{Proceedings of the IEEE/CVF international conference on computer vision}}. \bibinfo{pages}{3836--3847}.
	\newblock
	
	
	\bibitem[Zhang et~al\mbox{.}(2024)]%
	{zhang2024motiondiffuse}
	\bibfield{author}{\bibinfo{person}{Mingyuan Zhang}, \bibinfo{person}{Zhongang Cai}, \bibinfo{person}{Liang Pan}, \bibinfo{person}{Fangzhou Hong}, \bibinfo{person}{Xinying Guo}, \bibinfo{person}{Lei Yang}, {and} \bibinfo{person}{Ziwei Liu}.} \bibinfo{year}{2024}\natexlab{}.
	\newblock \showarticletitle{Motiondiffuse: Text-driven human motion generation with diffusion model}.
	\newblock \bibinfo{journal}{\emph{IEEE transactions on pattern analysis and machine intelligence}} \bibinfo{volume}{46}, \bibinfo{number}{6} (\bibinfo{year}{2024}), \bibinfo{pages}{4115--4128}.
	\newblock
	
	
	\bibitem[Zheng et~al\mbox{.}(2023)]%
	{zheng2023locally}
	\bibfield{author}{\bibinfo{person}{Xin-Yang Zheng}, \bibinfo{person}{Hao Pan}, \bibinfo{person}{Peng-Shuai Wang}, \bibinfo{person}{Xin Tong}, \bibinfo{person}{Yang Liu}, {and} \bibinfo{person}{Heung-Yeung Shum}.} \bibinfo{year}{2023}\natexlab{}.
	\newblock \showarticletitle{Locally attentional sdf diffusion for controllable 3d shape generation}.
	\newblock \bibinfo{journal}{\emph{ACM Transactions on Graphics (ToG)}} \bibinfo{volume}{42}, \bibinfo{number}{4} (\bibinfo{year}{2023}), \bibinfo{pages}{1--13}.
	\newblock
	
	
	\bibitem[Zhong et~al\mbox{.}(2025)]%
	{zhong2025sketch2anim}
	\bibfield{author}{\bibinfo{person}{Lei Zhong}, \bibinfo{person}{Chuan Guo}, \bibinfo{person}{Yiming Xie}, \bibinfo{person}{Jiawei Wang}, {and} \bibinfo{person}{Changjian Li}.} \bibinfo{year}{2025}\natexlab{}.
	\newblock \showarticletitle{Sketch2anim: Towards transferring sketch storyboards into 3d animation}.
	\newblock \bibinfo{journal}{\emph{ACM Transactions on Graphics (TOG)}} \bibinfo{volume}{44}, \bibinfo{number}{4} (\bibinfo{year}{2025}), \bibinfo{pages}{1--15}.
	\newblock
	
	
\end{thebibliography}

\end{document}